\documentclass[conference]{IEEEtran}
\usepackage{cite}

\ifCLASSINFOpdf
  \usepackage[pdftex]{graphicx}
  \graphicspath{{./img/}{./img/Samples/}{../pdf/}{../jpeg/}}
\else
  \usepackage[dvips]{graphicx}
  \graphicspath{{../eps/}}
\fi
\usepackage{color}

\usepackage[cmex10]{amsmath}
\usepackage[caption=false,font=footnotesize]{subfig}
\usepackage{url}

\usepackage{multirow}
\usepackage{bibentry}  

\begin{document}
%
\title{A Novel Algorithm for Real-time Procedural Generation of Building Floor Plans}

\author{\IEEEauthorblockN{Maysam Mirahmadi, Abdallah Shami}
\IEEEauthorblockA{Department of Electrical and Computer Engineering\\
The University of Western Ontario\\
London, Ontario\\
Email: \{mmirahma, ashami2\}@uwo.ca}
}


%


\maketitle

\begin{abstract}
Real-time generation of natural-looking floor plans is vital in games with dynamic environments. This paper presents an algorithm to generate suburban house floor plans in real-time. The algorithm is based on the work presented in \cite{Marson2010}. However, the corridor placement is redesigned to produce floor plans similar to real houses. Moreover, an optimization stage is added to find a corridor placement with the minimum used space, an approach that is designed to mimic the real-life practices to minimize the wasted spaces in the design. The results show very similar floor plans to the ones designed by an architect.
\end{abstract}



%
\IEEEpeerreviewmaketitle

\section{Introduction}
Dynamic generation of virtual environments gains popularity in modern games. Generating dynamic virtual environments for each round of the game allows even the savvy gamers to enjoy the game. A lot of games' scenes takes place in a town which has to be generated either manually or with a rule-based procedure. In some massive multiplayer games, producing the whole world is required. 

Besides the scale of the virtual environment, its details are also essential. Creating environments where the player can go inside the buildings adds a layer of design which multiplies its complexity. As a result, creating and managing such environments constitute a large portion of game design. 

A model describing the architecture of a city or inside the buildings can also be used in different disciplines. As an instance, such a description can be used to statistically model the signal propagation in built-up areas or used as a hypothetical signal propagation benchmark which can be customized to adapt to different scenarios. Our proposed model has been used in \cite{mirahmadiICC} to statistically model the propagation of a radio signal transmitted from an indoor transmitter to outside, which extremely depends on the floor plan of the building that encloses the transmitter.

In this paper, we introduce a novel procedural floor plan generation for suburban houses. The remaining of the paper is organized as follows. Section II reviews some of the previous work related to automatic generation of buildings. Section III describes the proposed algorithm. Section IV discusses the results, and finally Section V concludes the paper.

\section{Literature review}

Several algorithms have been proposed in literature to create architectures. The functionality of these algorithms can be categorized in three different classes: Creating the city maps, creating the 3-dimensional appearance of a city or a building, and creating the interior design of a building. 

M\"{u}ller \textit{et al.} propose a rule-based algorithm capable of generating streets and urban-looking buildings which results in realistic environments with high visual quality \cite{Muller2006}. Several simple 3D shapes, such as cubes, cones, etc, are placed or intersected together to form the shell of the buildings. Then the proposed shape grammar, which is an extensive set of rules governing the shapes, is processed to refine the buildings. First, the grammar looks for areas that shapes intersect. The algorithm discards these areas when putting windows or other ornaments. The next set of grammar rules tries to snap windows and other accessories to the visual lines in the structure which makes the resulting building more realistic. The algorithm produces high quality modern looking city landscape. However, the disadvantage is that it cannot be run in real-time. Running the algorithm possibly takes up to several days depending on the size of the requested scenario. In addition, it just creates only the shell of the buildings. Therefore, it has to be augmented if the interior design is required. 

The work presented in \cite{Hahn2006} focuses on real-time interior generation of large buildings, such as office buildings. The authors propose a method to manage the huge memory and processing power requirement for real-time floor plan generation of large buildings as well as maintaining the environment persistent. The building spaces are generated top to down. Therefore, it is possible to generate them in a lazy fashion and create the spaces that are necessary. If the player goes into a space, that space is then divided into smaller rooms and other accessories are added to it. The algorithm generates the floor plan in the following steps: building setup, floor division, Hallway division, room cluster division, and built region generation. In each step, more details are added to the floor plan, while the invisible architectures are removed form the cache. This approach is highly scalable and can be applied to scenarios with a number of large buildings without significant increase in memory or processing power requirements.

Another algorithm that is focused on creating the interior floor plan for buildings, especially for houses, is proposed in \cite{Marson2010}. The algorithm divides the available space using Squarified Treemap algorithm \cite{Bruls1999}. It then connects the rooms together and placed doors between them based on a connection graph which is randomly produced at a previous stage based on some rules. An example rule can be expressed as bedrooms should not be connected together. In the next step, if the rooms that are supposed to be connected are not adjacent, a procedure is applied to place a corridor to connect the rooms. However, the procedure wastes some space in the house which is not desirable and makes the floor plans unreal. 

Since the creation of the building appearance is highly required, the issue has been addressed and several academic and commercial tools exist to generate urban environment with great realism. However, to the best of our knowledge, there are just a few algorithms that can create building interior and none of them gives a detailed realistic floor plan and most often they just fit several rectangles which demonstrate rooms into a predefined area. 

\section{Proposed algorithm for constructing floor plans}
This section describes our algorithm for creating floor plans.
The proposed algorithm is based on the work presented in \cite{Marson2010}. However, the corridor placement stage is revised greatly and an optimization stage is also added to the design to make the resulting floor plans more realistic. Moreover, several rules have been introduced to prohibit the creation of bizarre looking rooms. 
Just like the main work, the algorithm has been adjusted to create random floor plans for suburban houses. However, it can be used to create the floor plan for any type of building with some modifications.

The proposed algorithm creates floor plans in the following steps:

\begin{enumerate}
\item Determining the area and outer shape of the house. 
\item Placing the rooms inside the house.
\item Creating the connectivity graph. This graph specifies if it is a way from one room to the other. 
\item Creating a corridor to connect the rooms together, if necessary.
\item Placing windows and doors.
\end{enumerate}
Each of the above steps is described in detail in the corresponding section.

The algorithm is primarily based on the following concepts.

\begin{itemize}
	\item The outer shape of the house, or its facade, is rectangular. 			
	\item Each house is composed of several connected rooms.
	\item No two spaces of the house can overlap.
	\item The building shape is preferred to be square or square-like. \textit{i.e.}, the square shape is preferred over long rectangles.
	\item No room can protrude the supposed boundary of the house.
	\item All significant portions of space should be used \cite{Hahn2006}.
	\item The area of each room is a random variable whose statistical properties are based on its functionality. 
	\item The floor plan should be connected. \textit{i.e.}, there is at least one way to go from one room to the other. In other words, the connectivity graph is a connected graph.
	\item Generating narrow long rooms which would be perceived unnatural should be avoided as much as possible.
	\item There are several windows, and at least a door connecting the house to the outside.
\end{itemize}

These notions are implemented in the proposed algorithm. The following sections explain the algorithm in detail.

\subsection{Determining the outer shape of the house}

The outer shape of most of the buildings are rectangular and does not have curved facade. The buildings with unusual shapes are in most cases public buildings with architectural importance. The proposed algorithm focuses on typical buildings and does not include these types of buildings. 

The shape of building facade is modeled by its aspect ratio, which is defined as,

\begin{equation}
AR = max(b/h, h/b)
\end{equation}
where $AR$, $b$, and $h$ are the aspect ratio, base and height of the rectangle representing the building facade, respectively. An area with a narrow shape, like a rectangle with large aspect ratio, is simply regarded as unsuitable for building in real life and thus, discarded in the proposed algorithm. 

\begin{figure}[t]
\centering
\includegraphics[width=.99\columnwidth]{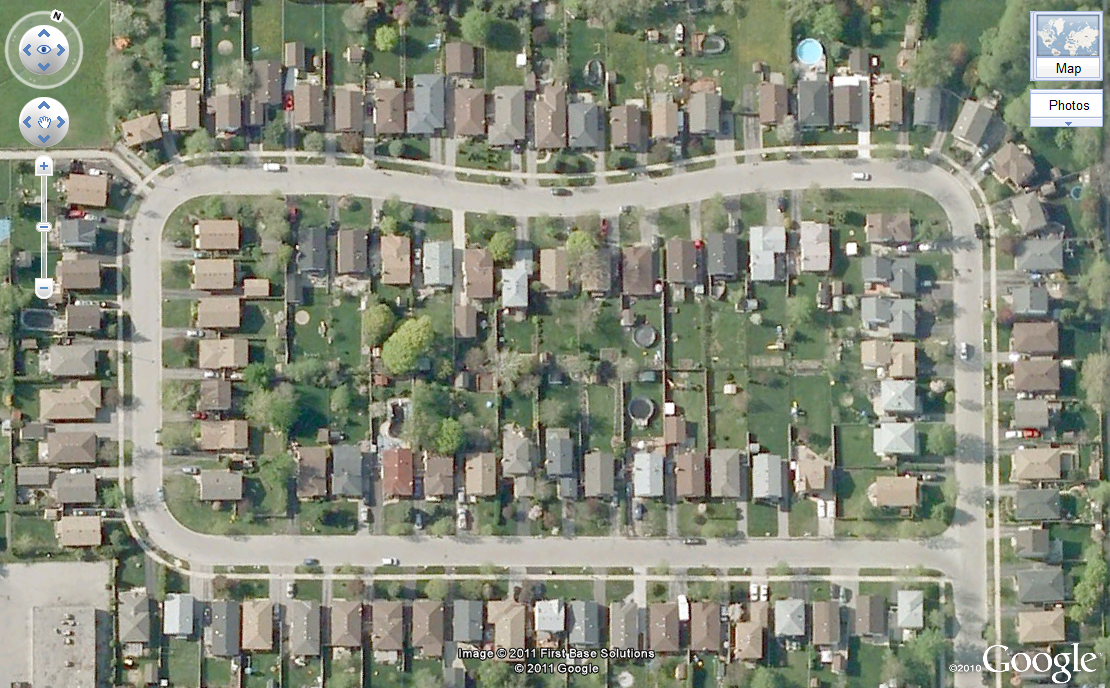}
\caption{An aerial photo showing the similarity of homes in a neighbourhood. [Courtesy of Google Inc.]}
\label{fig:arial}
\end{figure}

The area and aspect ratio of the buildings are random variables with predetermined distribution. For a given neighbourhood, the statistical properties of these random variables can be simply sampled from aerial photos. Fig. \ref{fig:arial} shows an aerial image of a residential region in London, Ontario. This type of suburban landscape is quite widespread in North America and it is probably the main design scheme of the residential areas. 

The output of the first step of the algorithm is a rectangle representing outer walls of the building. 

\subsection{Placing the rooms}

Each house is composed of rooms and probably corridors to connect them together. The number of rooms in a house varies from house to house, but some simple statistics can be found in census. The statistics such as the number of rooms and bedrooms are enough for our modeling purpose. Moreover, the parameters can be adjusted manually to simulate any virtual scenario.

In the proposed algorithm, the random variables are generated based on the joint probability distribution of the number of bedrooms and the number of rooms which is extracted from 2001 census of Canada \cite{census} and shown in Table \ref{tab:joint_prob}. It lets the generator to follow the real distribution of a neighbourhood.

\begin{table*}
\caption{Joint probability of the number of bedrooms and the number of rooms.}
\begin{tabular}{|l|*{10}c|}
\hline
Number of rooms $\rightarrow$ & 1 & 2 & 3 & 4 & 5 & 6 & 7 & 8 & 9 & 10 \\
\hline
no bedroom &  8.0338e-3 & 1.4385e-2 & 6.3392e-4  &          0  &          0     &       0   &         0  &         0    &        0    &        0\\
\hline
one bedroom &            0        &    0&  8.0221e-2 & 4.3408e-2&  1.1586e-2 & 3.4386e-3 & 1.1857e-3 & 5.8376e-4 & 2.1361e-4 & 1.0810e-4\\
\hline
two bedrooms &            0       &     0        &    0 & 9.6780e-2 & 9.0908e-2 & 3.9446e-2 & 1.6442e-2 & 8.5631e-3 & 2.9806e-3 & 1.5026e-3\\
\hline
three bedrooms &            0  &          0         &   0       &     0 & 7.6431e-2  & 1.0599e-1&  8.0343e-2 & 5.4633e-2 & 2.5820e-2&  2.5868e-2\\
\hline
four bedrooms &    0      &      0      &      0     &       0       &     0 &1.4412e-2 & 3.4647e-2&  5.2167e-2&  3.9863e-2 & 6.9406e-2\\
\hline

\end{tabular}

\label{tab:joint_prob}
\end{table*}

The area of each room is a random variable whose distribution depends on its functionality. As an instance, the living room is typically the largest room in a house, while the storage rooms are the smallest ones. According to \cite{Marson2010}, rooms of a house can be divided into three categories based on their functionality; service area includes kitchen and laundry room, private area composed of bedrooms and bathrooms, and social area like living room and dining room. After generating the area of each room, the area of each part of the house, \textit{i.e.}, social, service, and private part, is calculated. 

The rooms are then put into a hierarchical tree graph based on their functionality and the functionality of other rooms. The hierarchy starts from outside which is normally directly connected to the living room as the center of the daily life. The other rooms are included as branches below it. 

One important issue which is commonly left out in the algorithms is how to assign area to each room and how to determine if a house is equipped with a room. The proposed algorithm uses a collective distribution for all types of the rooms. This approach uses the census data to find some distribution parameters. 

To associate the functionalities to the generated rooms, the algorithm uses a list called priority list. It simply picks the $N$ most important rooms, where $N$ is the number of rooms in the house. Then, the area of each room, is specified based on the selected functionality which determines its distribution. To customize the output and match it to a different type of building, the priority list and area distributions can be adjusted.

\begin{figure}[t]
\centering
\includegraphics[width=.99\columnwidth]{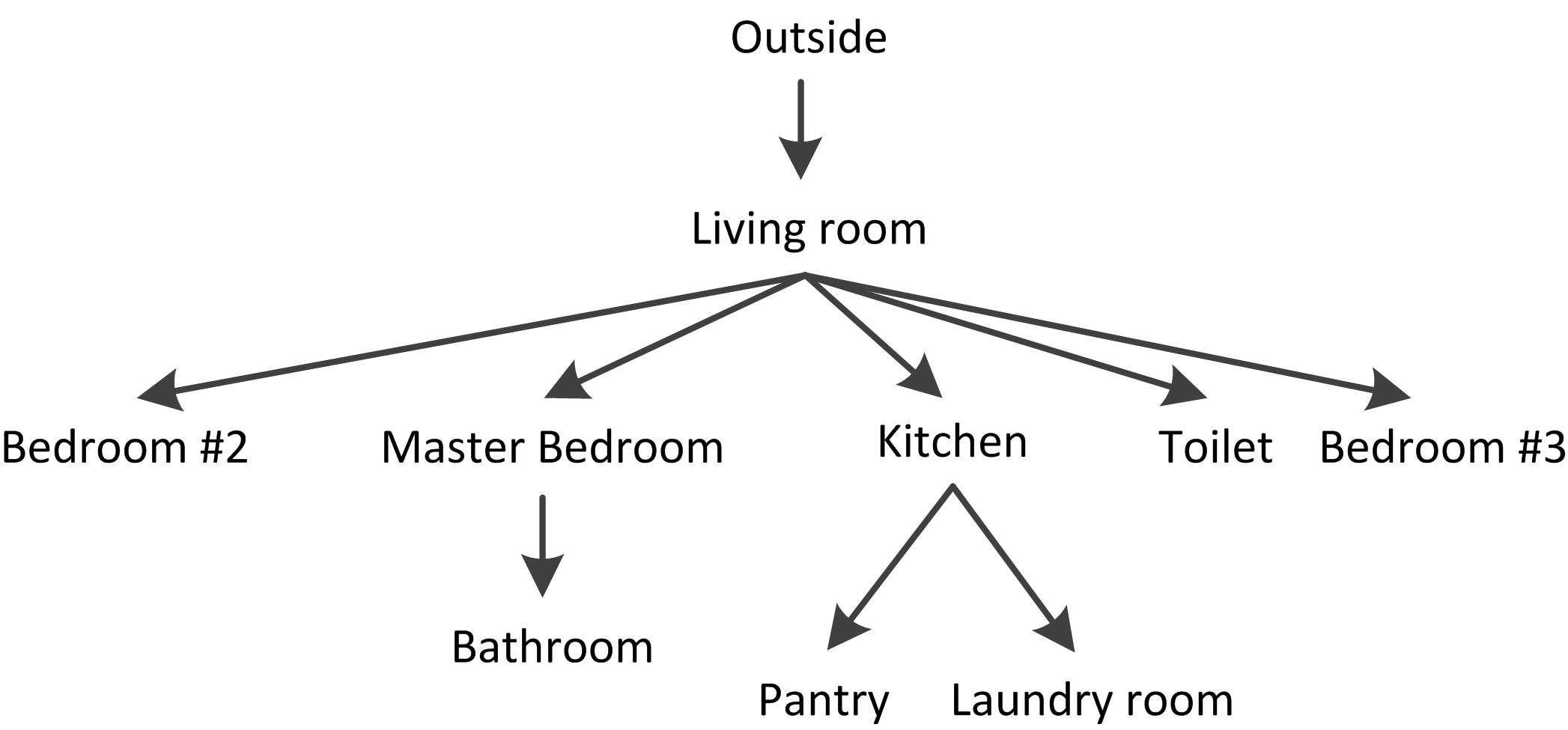}
\caption{Hierarchy graph of a sample house.}
\label{fig:room_hierarchy}
\end{figure}

In the next stage, a rule-based algorithm puts the rooms in the hierarchy tree based on their functionality. As an instance, the kitchen is connected to the living room either directly or via the dining room and hence, it is placed under the living room directly or via the dining room. In some cases, the position of a room in the hierarchy tree also depends on the functionality of other rooms. For example, in a typical house there should be a bathroom connected to the common area. The extra bathrooms are typically inside master bedroom or other large bedrooms. Therefore, there is a bathroom connected directly to the common area and there may be several other bathrooms connected to bedrooms. Some basic rules are as follows,
\begin{enumerate}
	\item Place outside node as the root.
	\item Place living room below outside.
	\item If there is a kitchen, place it below living room.
	\item If there is any bedroom, place the largest one under living room and name it Master bedroom.
	\item If there is just one bathroom, place it below living room.
	\item Place the remaining bathroom below bedrooms, starting from the largest bedroom.
	\item place laundry and pantry below kitchen, if any.
\end{enumerate}
The hierarchical tree of rooms of a sample house is demonstrated in Fig. \ref{fig:room_hierarchy}.

\begin{figure}
\centering
  \subfloat{
  \includegraphics[width=.45\columnwidth]{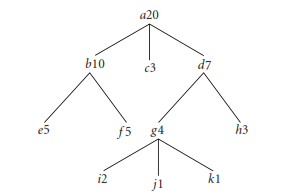} }
  \subfloat{
  \includegraphics[width=.45\columnwidth]{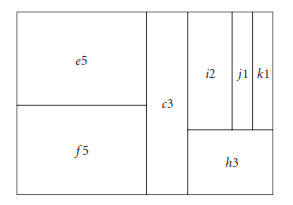} }
\caption{A tree graph used by Treemap algorithm and its corresponding rectangle division.}
\label{fig:treemap_tree_area}  
\end{figure}

The next step is placing the rooms in the house. An algorithm called Squarified Treemap \cite{Bruls1999}, places the rooms. The objective of the Squarified Treemap algorithm is to divide a region of space into several smaller regions with predefined area without any unused space. It also tries to minimize the aspect ratio of each block to be more square. Squarified treemap \cite{Bruls1999} algorithm is an extension of standard Treemap algorithm \cite{Johnson1991} which gives priority to the square or square-like shapes. The original Treemap algorithm organizes the spaces in a tree graph like the one shown in Fig. \ref{fig:treemap_tree_area}. As can be seen in the figure, the method can generate elongated rectangular subdivisions. These subdivisions are not favourable in floor plans. Therefore, Squarified treemap algorithm is adapted for the proposed algorithm. For a detailed explanation about Squarified Treemap algorithm refer to \cite{Bruls1999}.

\begin{figure}
\centering
  \subfloat [Step 1]{ 
  \includegraphics[width=.45\columnwidth]{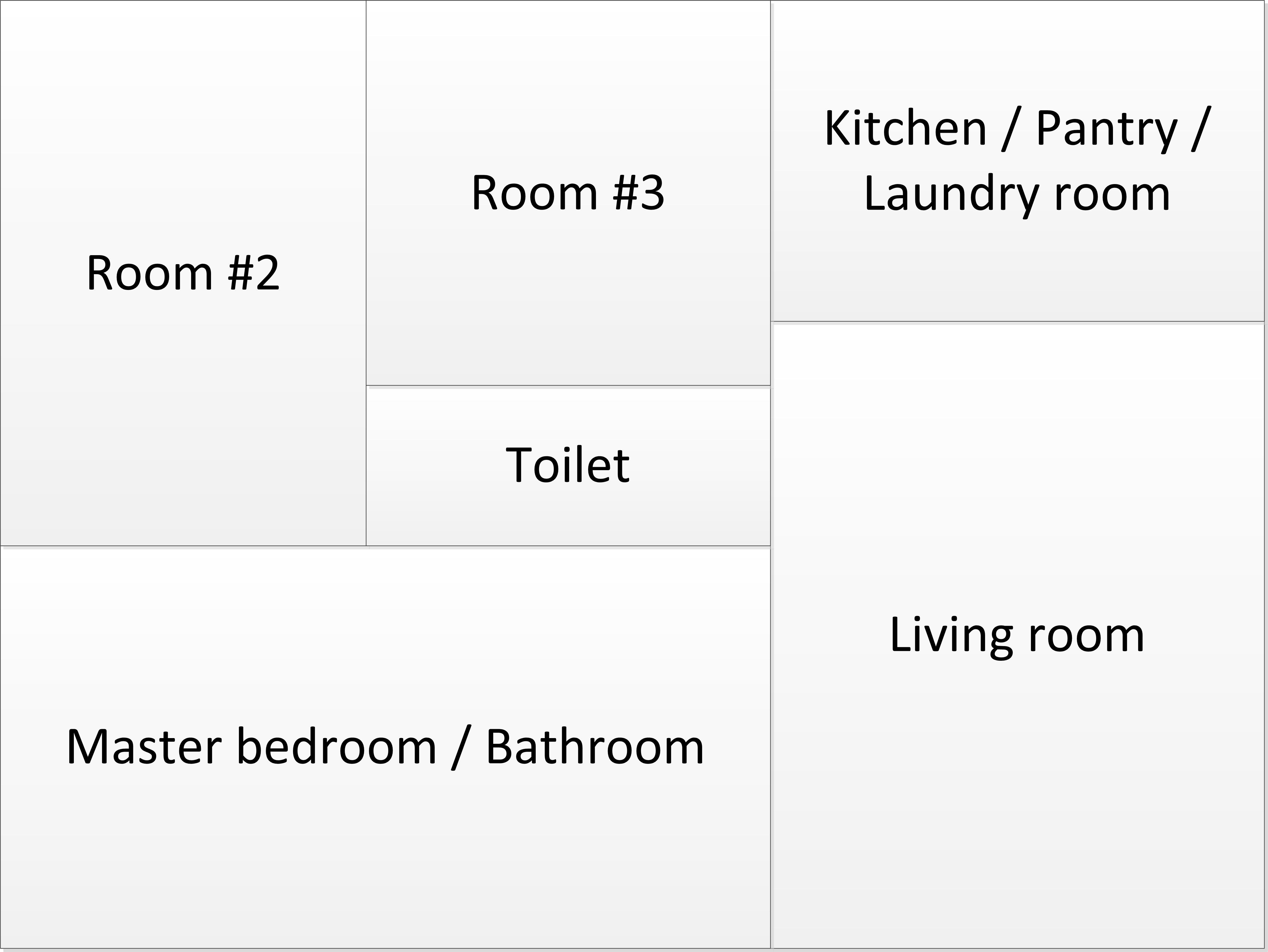} }
  \subfloat [Step 2] { 
  \includegraphics[width=.45\columnwidth]{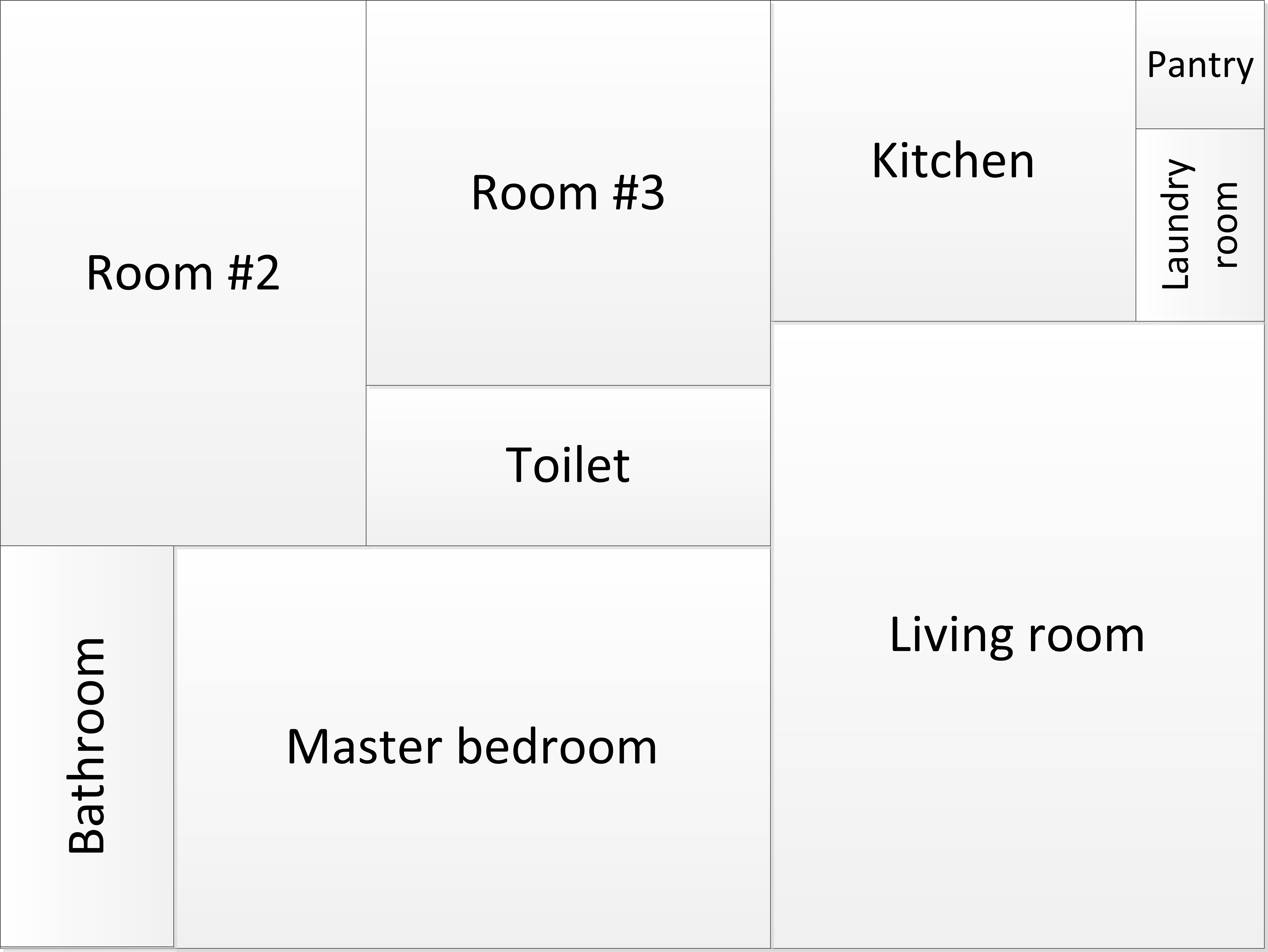} }
\caption{Room placement steps for the hierarchy tree shown in Fig. \ref{fig:room_hierarchy}.}  
\label{fig:roomPlacement}  
\end{figure}

The automatic floor plan generator algorithm uses Squarified Treemap in a step by step approach. It first places the rooms in the first level of the hierarchy. At this step, the algorithm puts rooms with total surface area of all the rooms below it. The it moves to each room and places the smaller rooms below it inside. The steps of putting rooms inside a house is illustrated in Fig. \ref{fig:roomPlacement}.

\begin{figure*}[t]
\centering
  \subfloat []{
  \includegraphics[width=.25\textwidth]{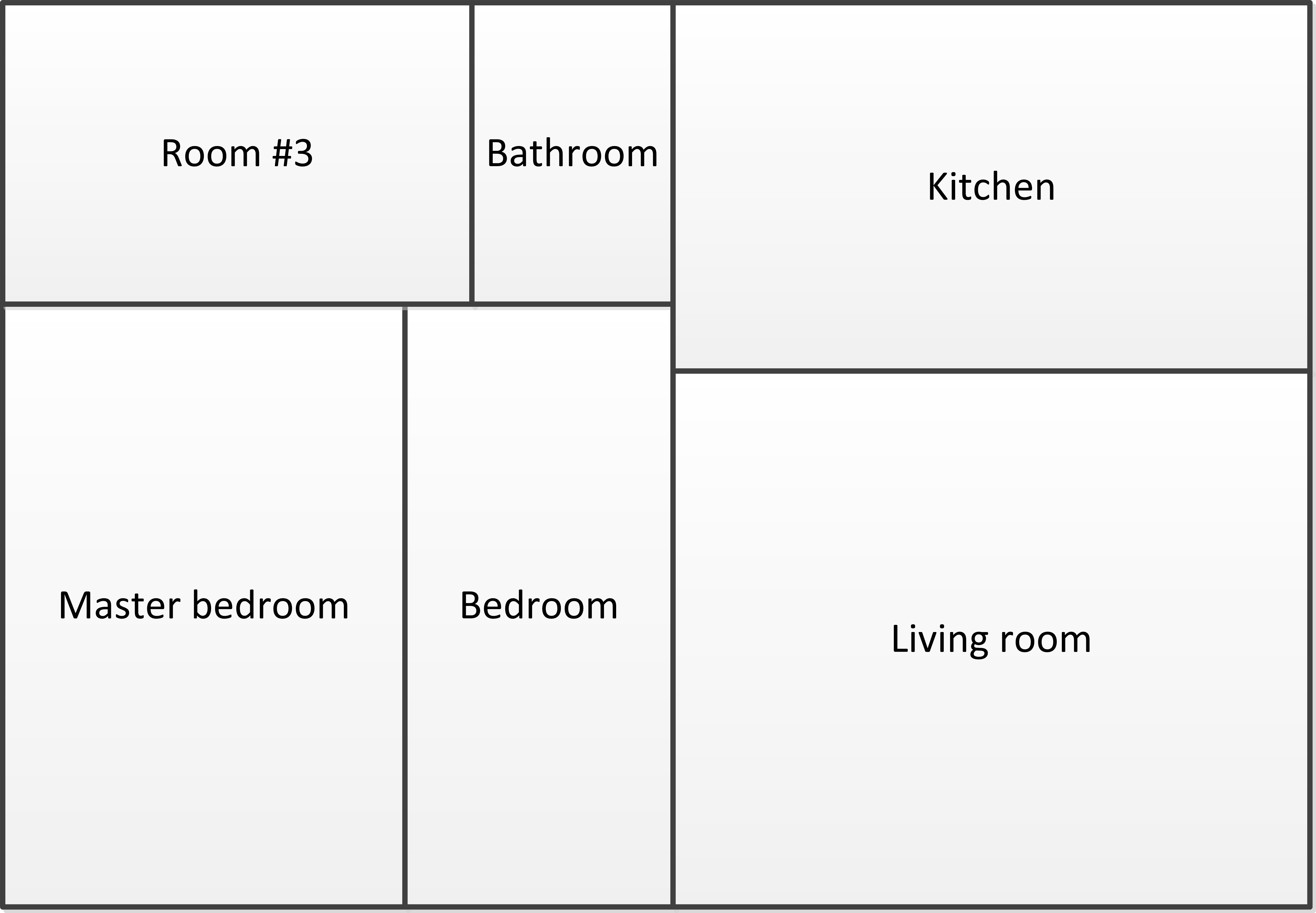}
  \label{fig:Corr1} }
\hspace{.05\textwidth}
  \subfloat []{
  \includegraphics[width=.25\textwidth]{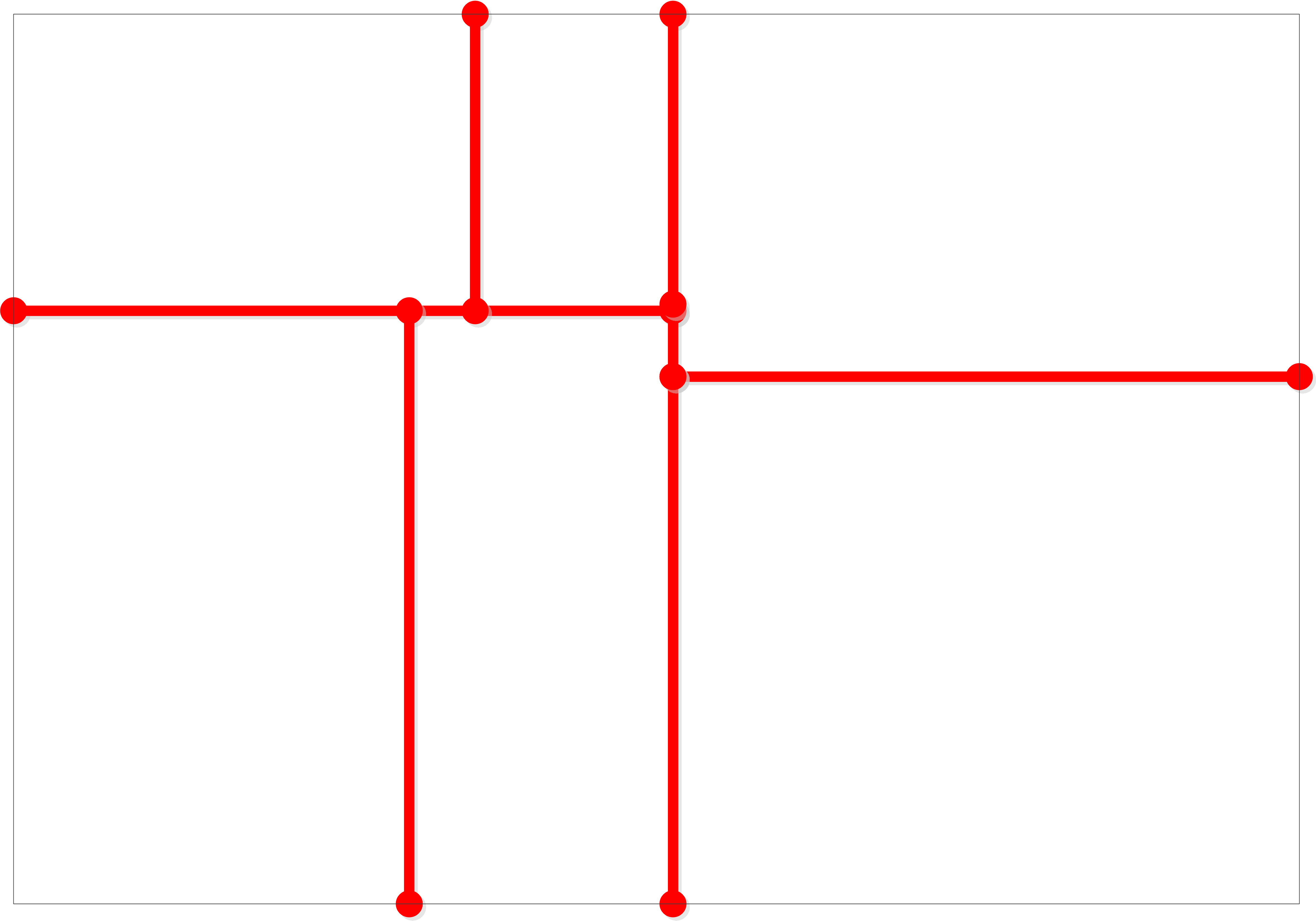}
  \label{fig:Corr2} }
\hspace{.05\textwidth}
  \subfloat []{
  \includegraphics[width=.25\textwidth]{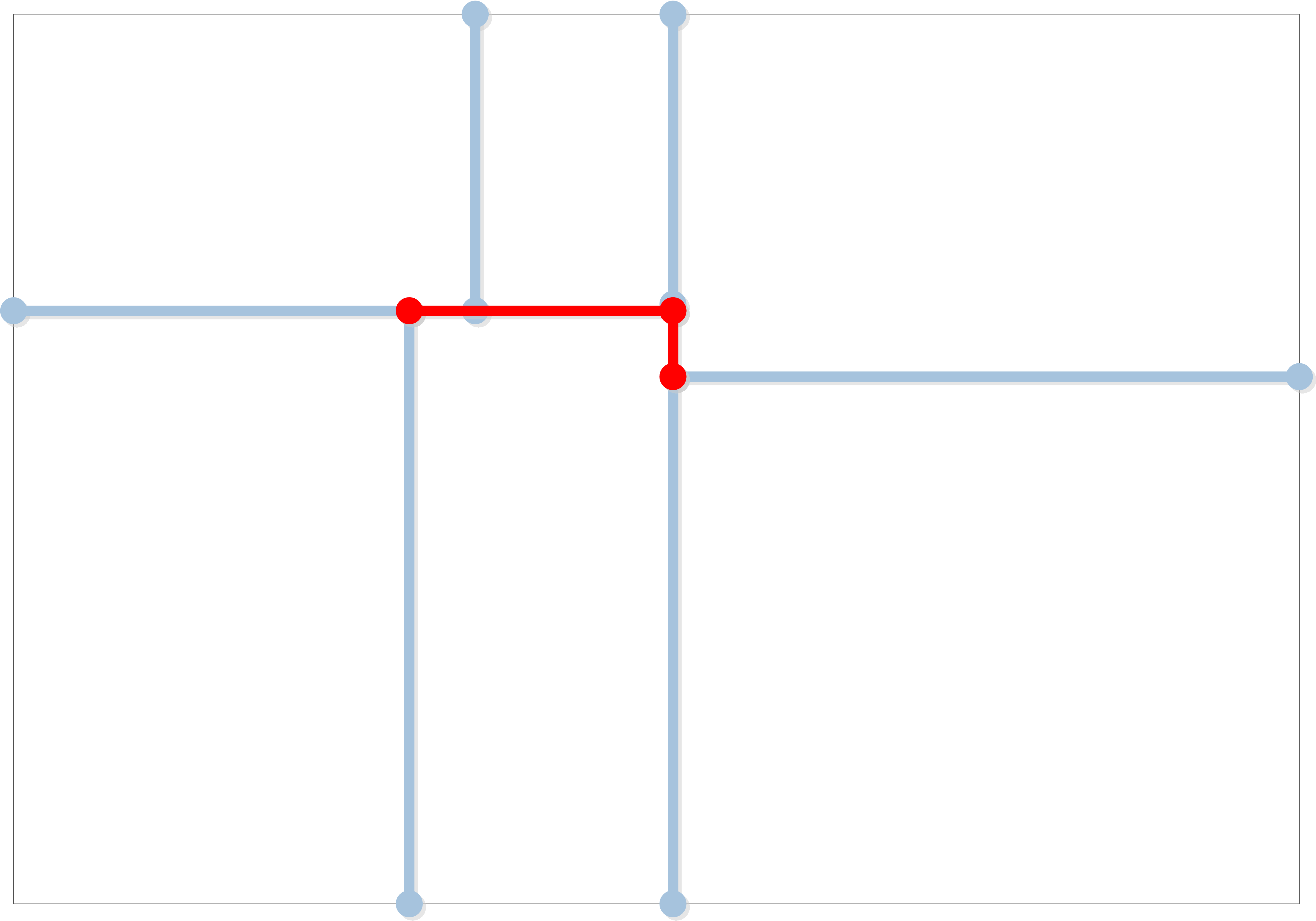}
  \label{fig:Corr3} }
\label{fig:Corr}
\caption{Finding the corridor path.}
\end{figure*}

\subsection{Corridor Placement}
\label{sec:corridor_placement}

In a typical house all of the rooms are connected either directly or via some corridors to the living room, as the center of activities. The proposed algorithm includes this fact into the floor plan designs. It identifies the rooms that are required to be connected directly to the living room and if they are not adjacent to the living room, it places a corridor, connecting the rooms to the living room.

Depending on the number and type of the rooms and their placement, it may be required to place a corridor to connect the rooms together. For example, in Fig. \ref{fig:Corr1} Room \#3 cannot be connected to the living room. Thus, a corridor is required to connect the leftmost room (Room \#3)  to the living room.

The Corridor Placement algorithm functions as follow:
\begin{enumerate}

\item The rooms that requires connections are identified. We refer to these rooms as the corridor rooms.
\item A graph is constructed with the walls of the corridor rooms as well as the living room. The outer walls of the building is not considered in this graph. (Figure \ref{fig:Corr2})
\item At the next stage the graph is pruned and the edges that connects to a vertex with degree one is removed from the graph. This graph is called corridor graph. (Figure \ref{fig:Corr3})
\item The shortest path in the graph connecting all the rooms in the graph is chosen using standard shortest path algorithm \cite{dijkstra1959note}. A room is considered connected if it shares an edge or a vertex with the graph. The use of the shortest path algorithm is justified by the fact that corridor area is a wasted area in the house and has to be minimized. Thus, we choose the shortest path which is translated to the smallest area.
\item If a room is connected by a vertex and does not have any shared edge, it is required to modify the graph to be able to place a door for the room. This is done through either shifting the edge of the graph inside the room or lengthen the edge connected to the shared vertex. The process is illustrated in Fig. \ref{fig:ModifyingCorr}.

All of the possible choices to modify the corridor graph constitutes the action set.
\begin{multline}
A = \{ a_i = {Shift | Lenghten} \}, \\ \forall e_i \in Edges, i = 1\dots Number\ of\ edges 
\end{multline}
Selecting proper actions for each edge results in a small corridor. Therefore, an optimization is required to prohibit the generation of bulky corridors and minimize corridor area. 

\begin{figure*}[ht]
\centering
\subfloat[Shifting the corridor downward to align with the living room wall]{
\includegraphics[width=.25\textwidth]{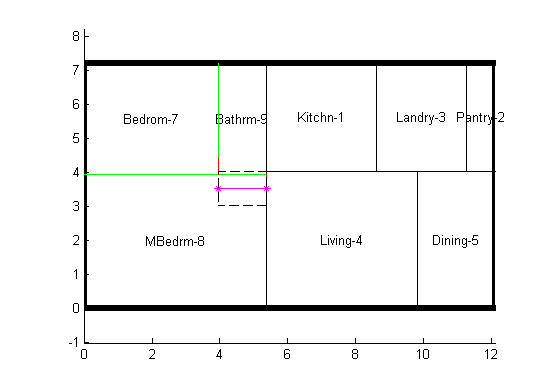}
\label{fig:ModifyingCorr1}
}
\hspace{.05\textwidth}
\subfloat[Lengthen the corridor to make room for a door to the living room]{
\includegraphics[width=.25\textwidth]{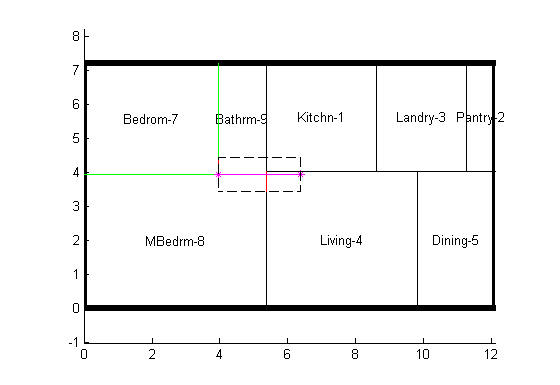}
\label{fig:ModifyingCorr2}
}
\hspace{.05\textwidth}
\subfloat[A combination of shifting upward and lengthen the corridor]{
\includegraphics[width=.25\textwidth]{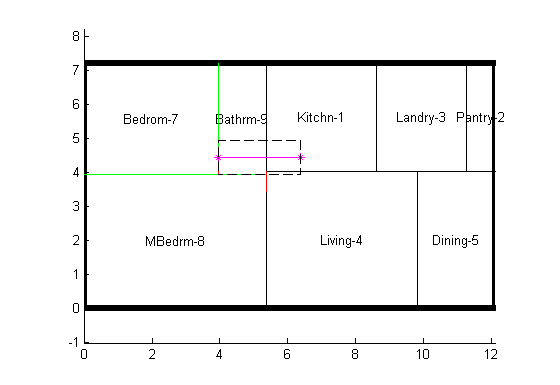}
\label{fig:ModifyingCorr3}
}
\label{fig:ModifyingCorr}
\caption{Optimizing the corridor with different actions.}
\end{figure*}

\item The corridor should not change the shape of surrounding rooms in a way that their space become unusable. Therefore, the corridor graphs that leaves peculiar shape rooms behind are filtered out. Then, An algorithm compares the area of the corridor resulting from each modified corridor graph and chooses the corridor with the smallest area. The resulting graph is called the optimized corridor graph.

\item At the last step, the polygon representing the corridor is constructed based on the optimized corridor graph and extruded from the overlapping rooms. The output is the final floor plan. It should be noted that the area of the corridor itself is regarded as the extension of the living room. Therefore, in all subsequent steps, the corridor walls are considered as the living room walls.

\end{enumerate}

\subsection{Placing Windows and Doors}

The next step places the connections between the rooms, \textit{i.e.}, doors, and windows in the floor plan. The studies \cite{Marson2010} shows that not all of adjacent rooms can be connected together. As an instance, bedrooms can not be connected to the kitchen. 

The hierarchy tree is normally the basis of connection graph which may then augmented by several other edges. For example, it is possible to connect kitchen and dining room. The decision to add these optional edges is taken randomly based on the functionality of both rooms. 

The doors are then placed randomly at the shared walls between the rooms. The door size is fixed and is adjusted manually by the algorithm designer. However, their position in walls are randomly chosen. 

The same approach is taken for placing windows. The only difference is that the connection graph is constructed regarding the placement of the rooms as well as some restrictive rules. As a general rule, the rooms that share a wall with outside are equipped with a window unless it is prohibited. For example, a sample rule is that a window cannot be installed in bathrooms.

\section{Results}
In this section, several generated floor plans as well as their parameters are presented. The parameters used to generate the floor plans are specified in Table \ref{tab:param}. 

It should be noted that the number of bedrooms and the number of rooms are random variables generated based on the distribution in Table \ref{tab:joint_prob}.

\begin{table}
\centering
\caption{List of the parameters used in generating floor plans.}
\begin{tabular}{|l|l|}
\hline
Parameter & Distribution \\ \hline
bedroom area & Uniform(8,18) \\
other room's area & Uniform(3,11) \\
building aspect ratio & Uniform(1,2)\\
\hline
\end{tabular}
\label{tab:param}
\end{table}

\begin{figure}[t]
\centering
\includegraphics[width=.9\columnwidth]{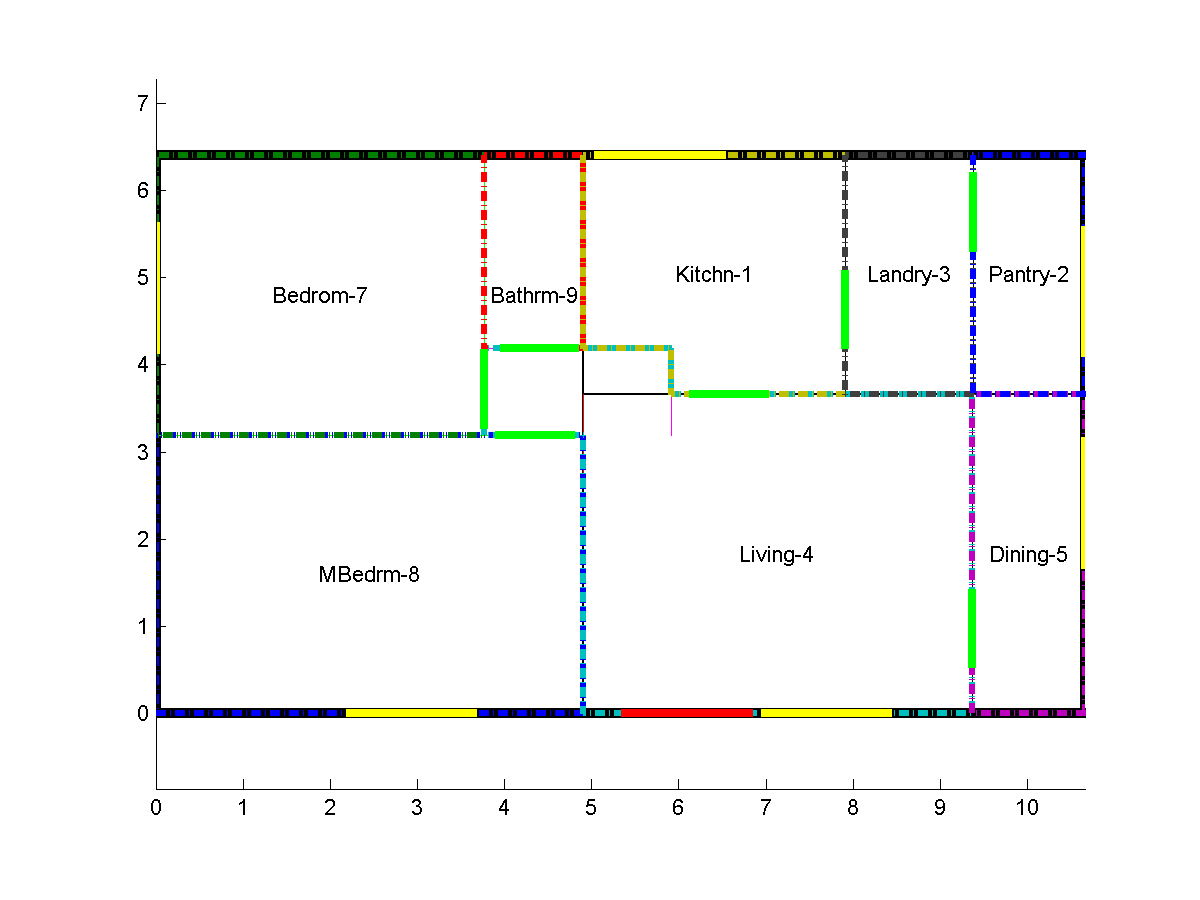}
\caption{A sample floor plan automatically generated by the algorithm.}
\label{fig:sample_floorplan}
\end{figure}

A sample floor plan generated with the parameters in Table \ref{tab:param} is shown in Fig. \ref{fig:sample_floorplan}. The house has two bedrooms and a bathroom. The room boundaries before adding corridor is depicted with thin lines whereas the dashed bold lines show the boundaries after placing the corridor. The floor plan looks natural and it could be used in any application. For example, it can be used by a game engine to produce authentic floor plans. The secondary bedroom was only connected to the master bedroom and bathroom. Since connecting two bedrooms together or through bathroom is prohibited in the rules, a corridor is placed to connect the secondary bedroom to the living room. The position of the corridor is optimized to occupy the least possible space.

Several other possibilities of the placing the corridor is shown in Fig. \ref{fig:ModifyingCorr}. Fig. \ref{fig:ModifyingCorr1} shows the case where the graph is shifted downward to align with living room walls, however, this case is not valid as the corridor can not accommodate a door to the smaller bedroom. Fig. \ref{fig:ModifyingCorr2} shows the case where a graph edge is lengthen to make room for a door from corridor to the living room. A combination of shifting upward and lengthening the corridor is shown in Fig. \ref{fig:ModifyingCorr3}. And in the final floor plan shown in Fig. \ref{fig:sample_floorplan}, the corridor generated by a combination of shifting upward to align with the room boundaries and lengthening to make room for a connection to the living room is selected. It is worth mentioning that according to the discussions in Section \ref{sec:corridor_placement}, the corridor with least consumed space is selected.

\begin{figure*}
\centering
  \subfloat[A generated floor plan with augmented details.] {
  \includegraphics[width=.45\textwidth]{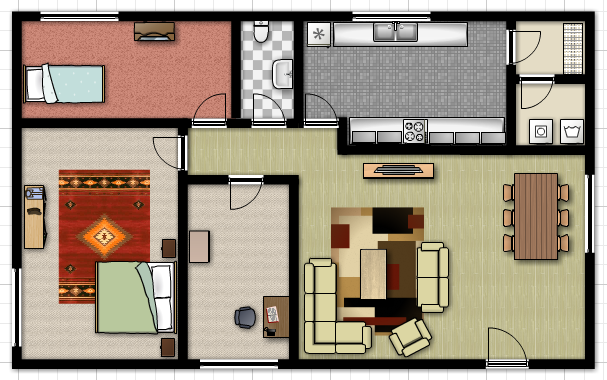}
  \label{fig:furnished}
  }
  \hspace{.05\textwidth}
  \subfloat[A real floor plan \cite{Marson2010}.] {
  \includegraphics[width=.45\textwidth]{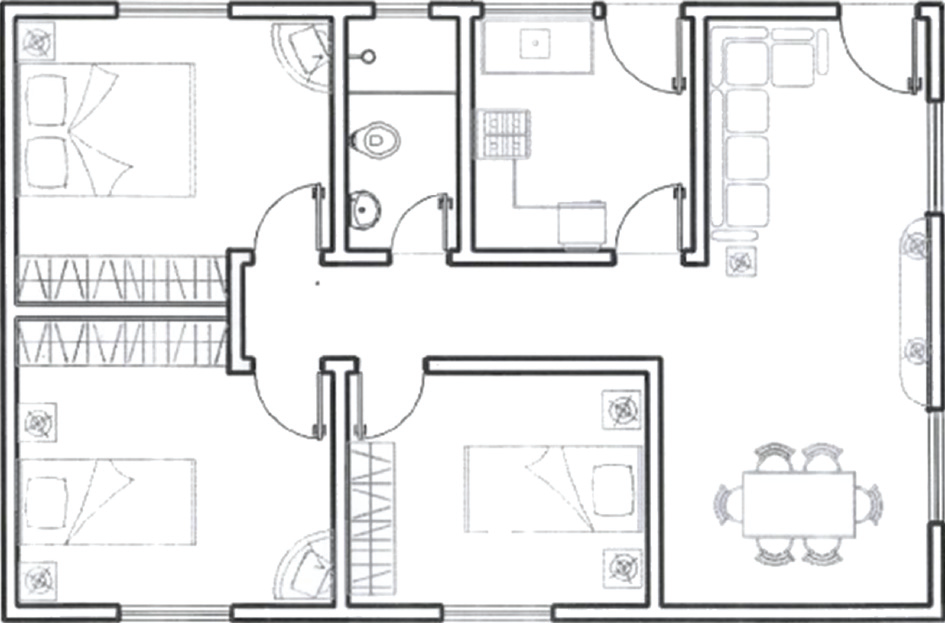}
  \label{fig:real_floorplan}
  }
  
  \subfloat[3D realization of the generated floor plan \cite{floorplanner}.] {
  \includegraphics[width=.45\textwidth]{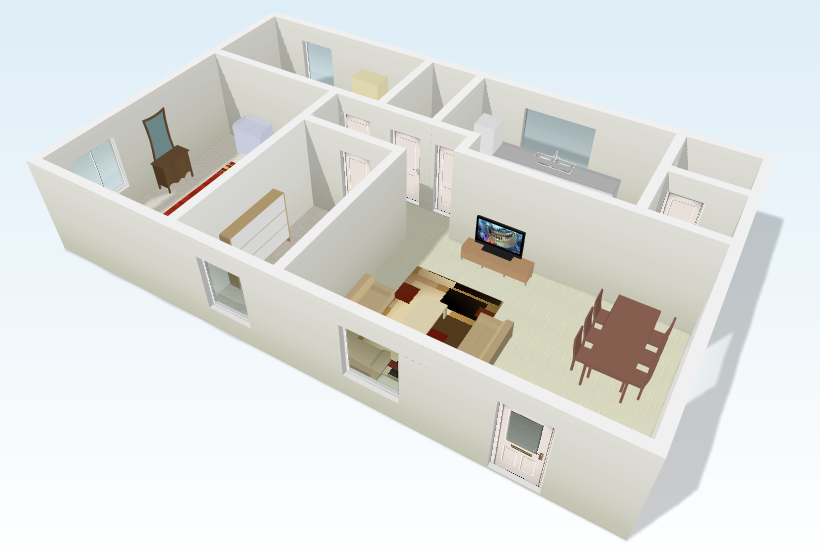}
  \label{fig:furnished_3d}  
  }
\end{figure*}

To give a sense of the generated floor plans, we have added a sample floor plan with some furnitures. The floor plan as well as its 3D realization is illustrated in Fig. \ref{fig:furnished} and \ref{fig:furnished_3d}. A real architect-designed floor plan is also shown in Fig. \ref{fig:real_floorplan} for the sake of comparison. The generated floor plan shows utmost similarity to the real one. It is worth mentioning that the improvements in the algorithm compared to the algorithm in \cite{Marson2010} prohibits the generation of long unused corridors which would make unused space. 

\section{Conclusion}
Dynamic generation of floor plan is highly required by games that generates the environment to provide the savvy gamers with the ultimate experience. Such a model can also be used in all of the problems concerning buildings. An effort to use the proposed model to predict the strength of a radio signal of an indoor transmitter, like a WiFi access point, outside the building has been reported in \cite{mirahmadiICC}. 

This work proposed an algorithm for real-time generation of floor plans. The algorithm focused on suburban houses. However, it can be tailored to accommodate other types of buildings. The algorithm generates every aspect of the house randomly, resulted in dissimilar floor plans. The random generation includes the outer shape of the house, its area, number of rooms and their functionality and the position of windows and doors.

An especial effort has been devoted to placing a corridor inside the house. The proposed algorithm optimizes the place and even shape of the corridor to produce natural-looking designs and reduce the area of the corridor which is regarded as a wasted space in house. 

The results show floor plans with utmost similarity to real floor plans. It can be imagined that the gamer that is playing inside the 3D realization of such generated floor plan, would find it very natural.




\bibliographystyle{IEEEtran}
\bibliography{Ref}
%

\clearpage
\onecolumn
\section*{Appendix A}
Several automated floor plans are illustrated in the following figures to show the resemblance to real floor plans and also their variety.

\begin{tabular}{|c|c|c|}
\hline 
\includegraphics[scale=.25]{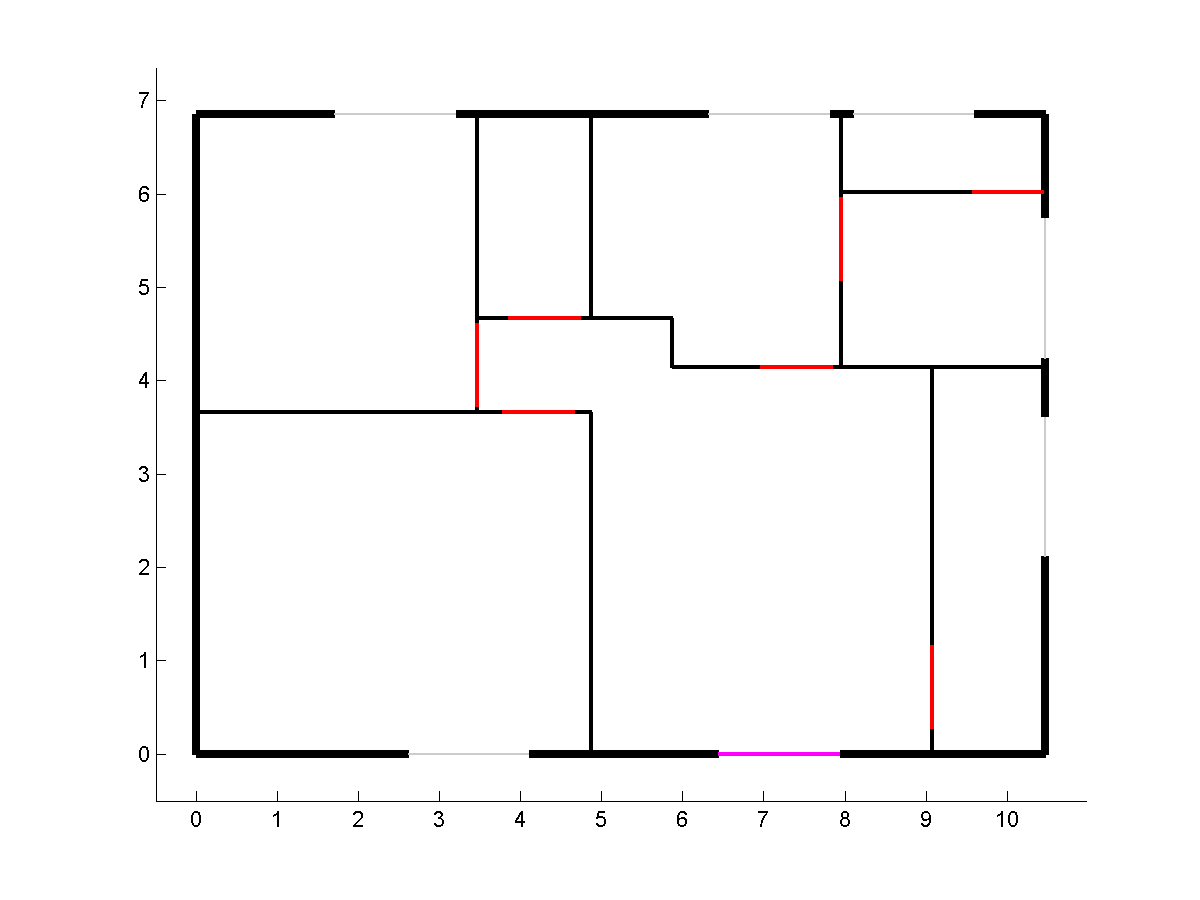} &  \includegraphics[scale=.25]{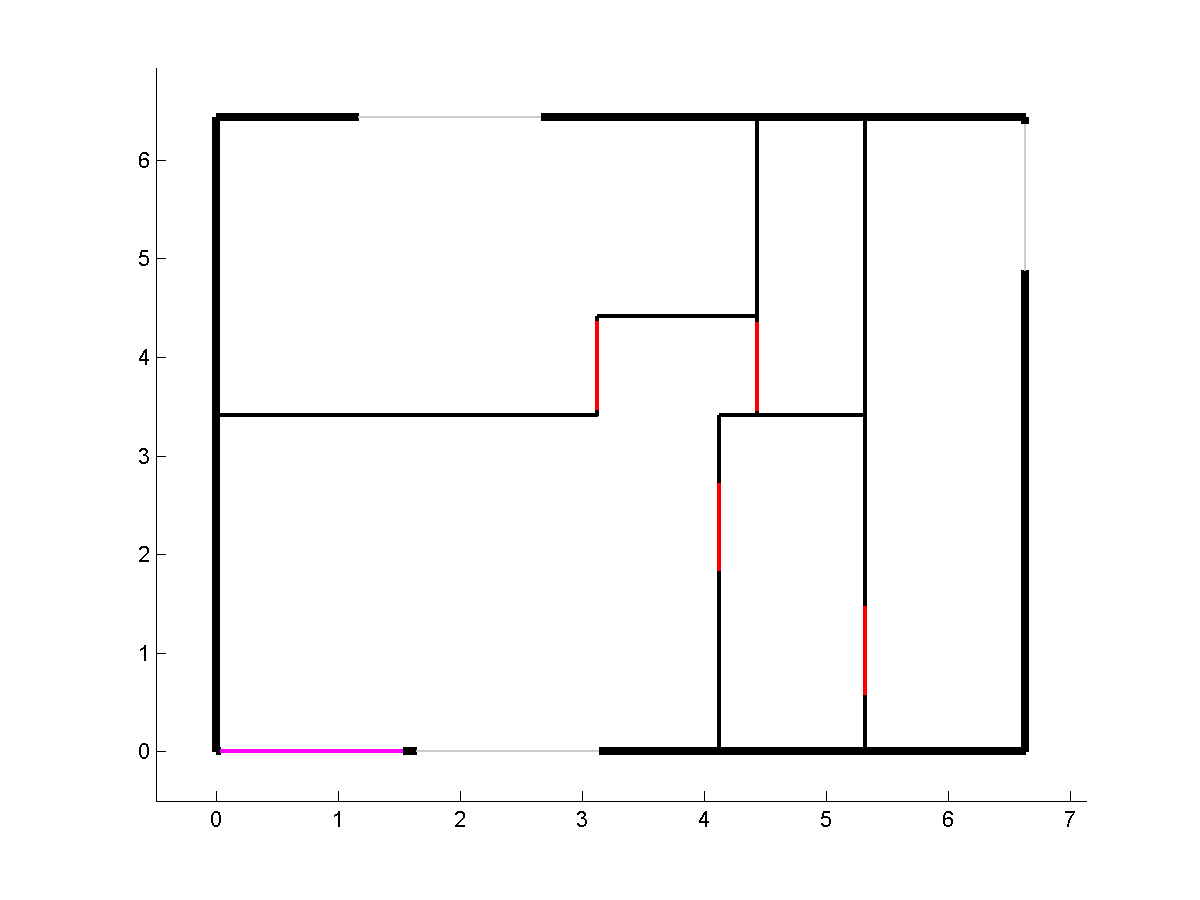} & \includegraphics[scale=.25]{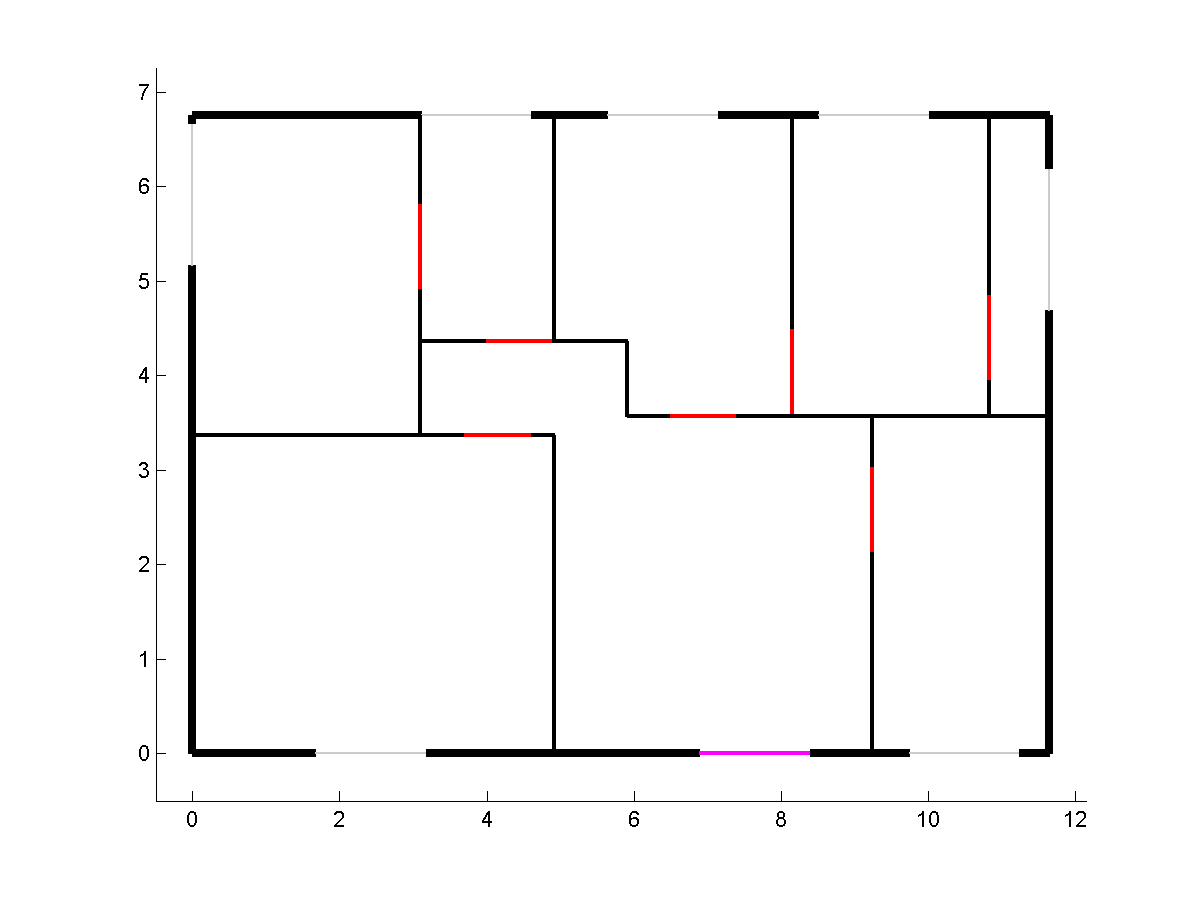}\\ 
\hline
\includegraphics[scale=.25]{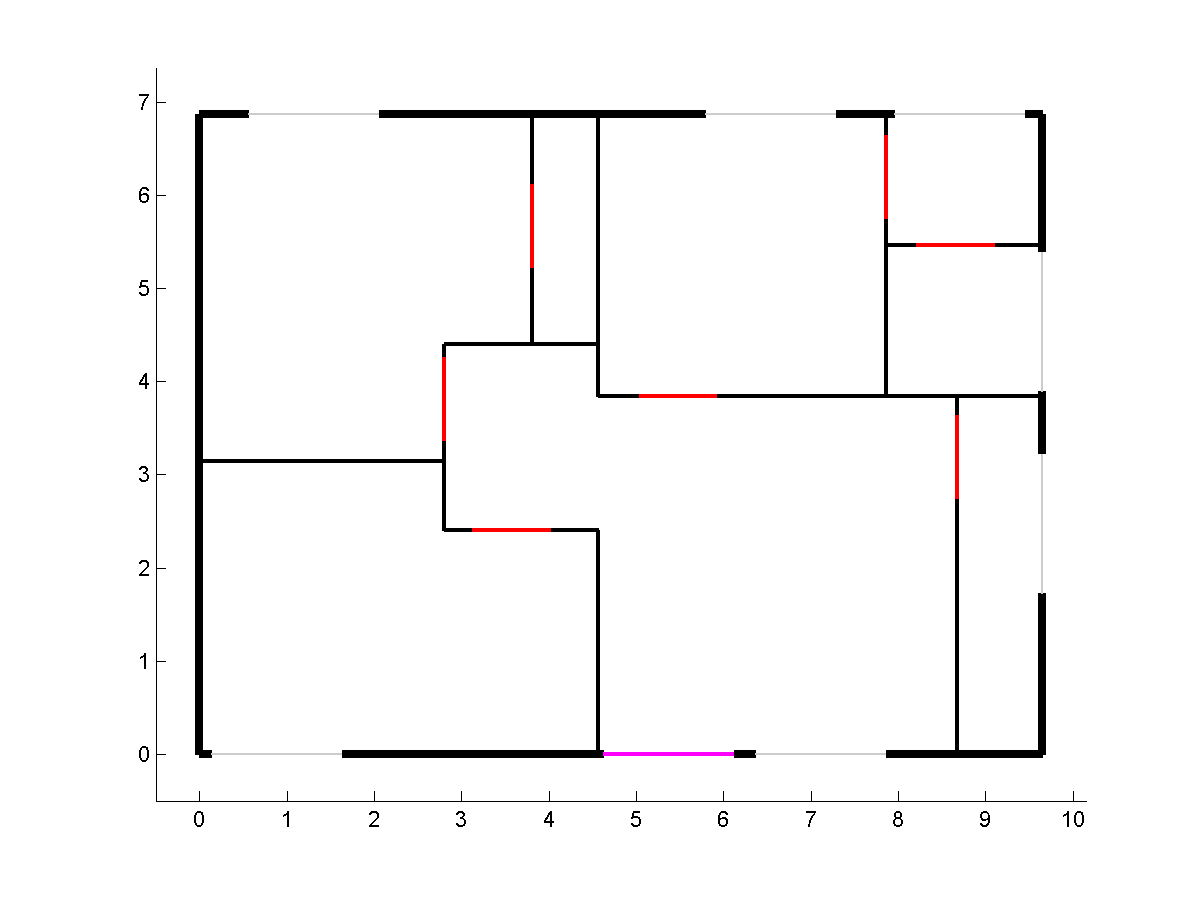} &  \includegraphics[scale=.25]{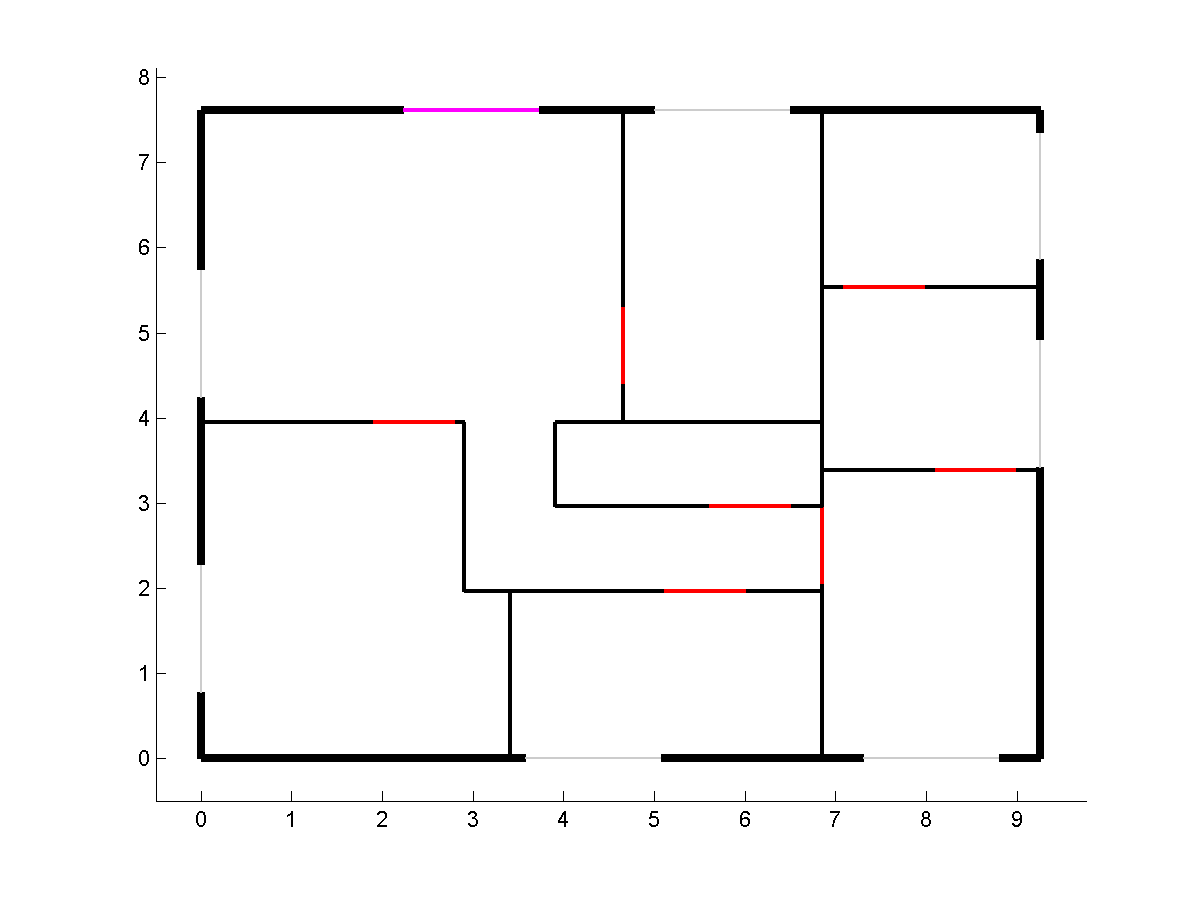} & \includegraphics[scale=.25]{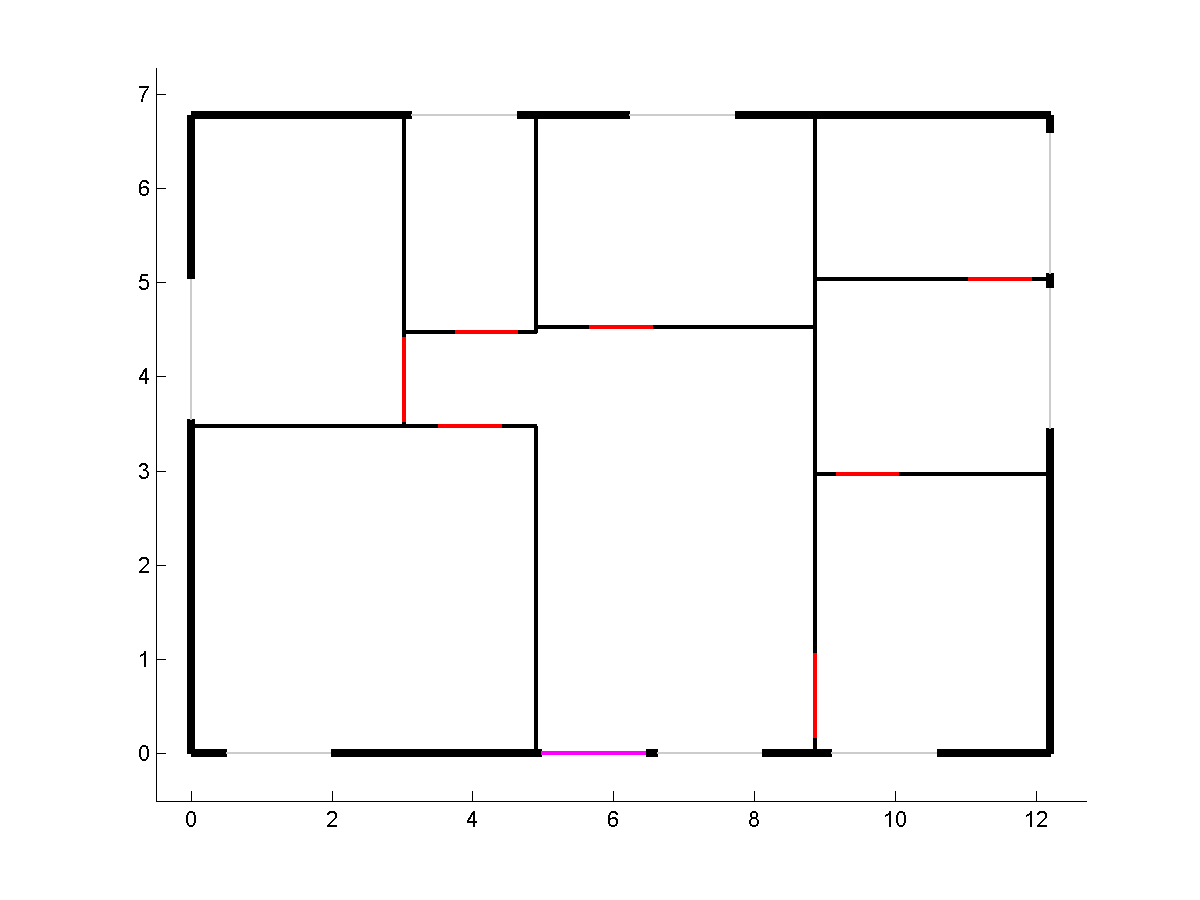} \\ 
\hline 
\includegraphics[scale=.25]{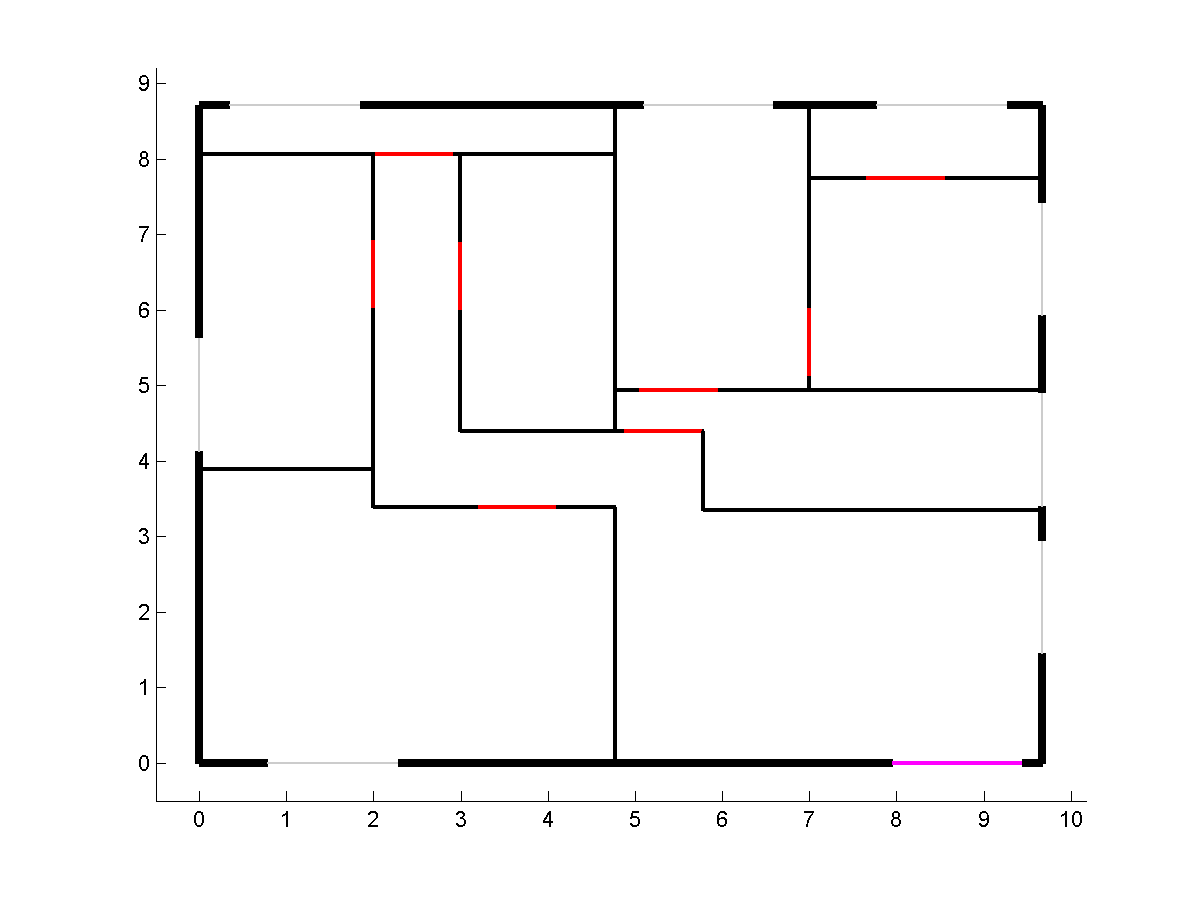} &  \includegraphics[scale=.25]{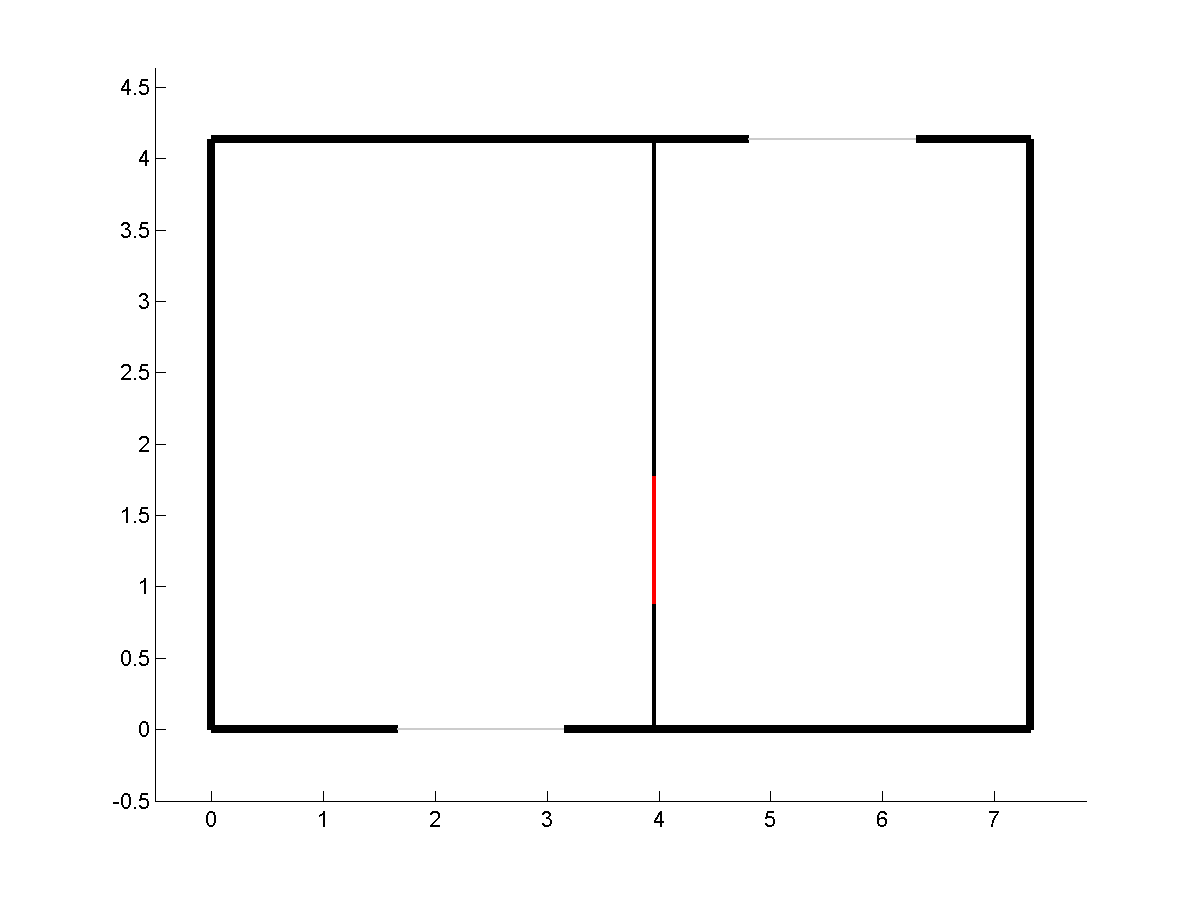} & \includegraphics[scale=.25]{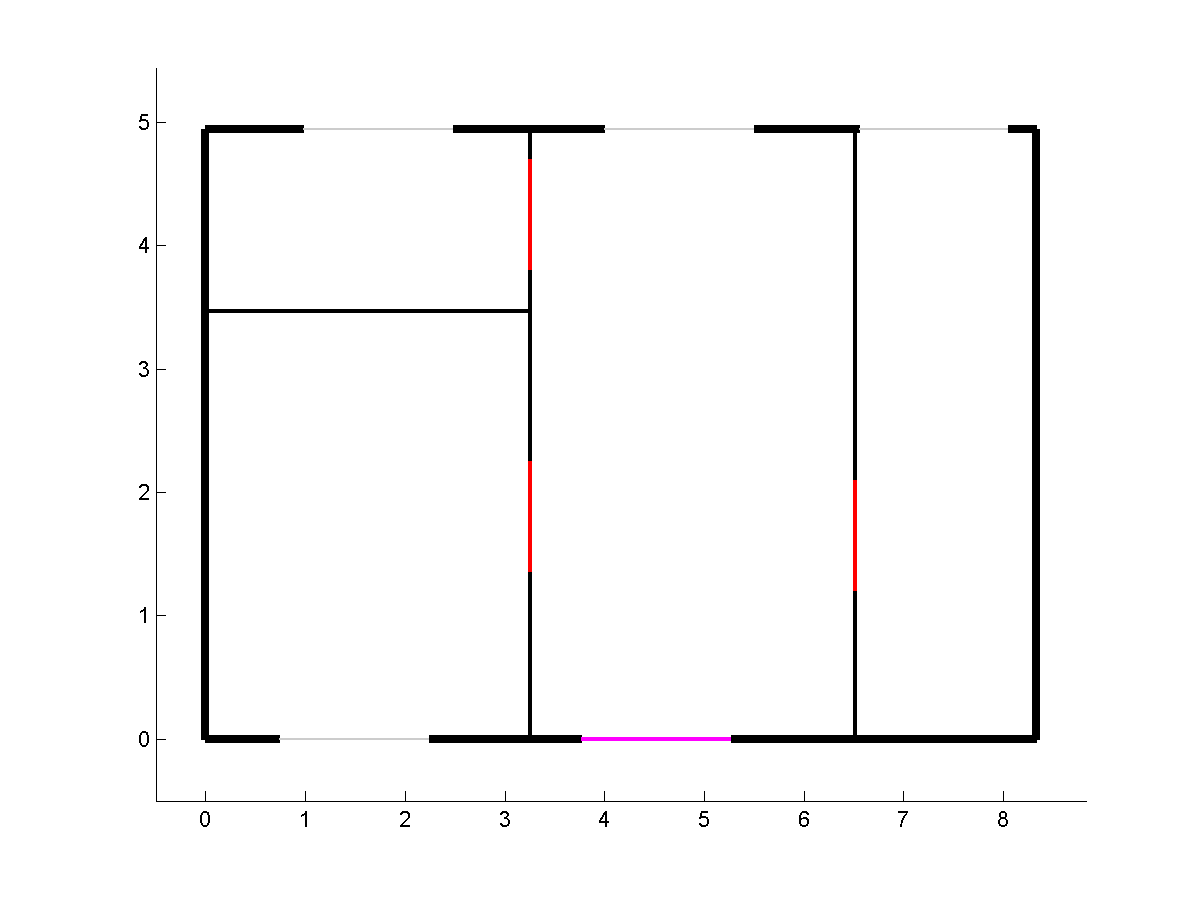} \\ 
\hline 
\includegraphics[scale=.25]{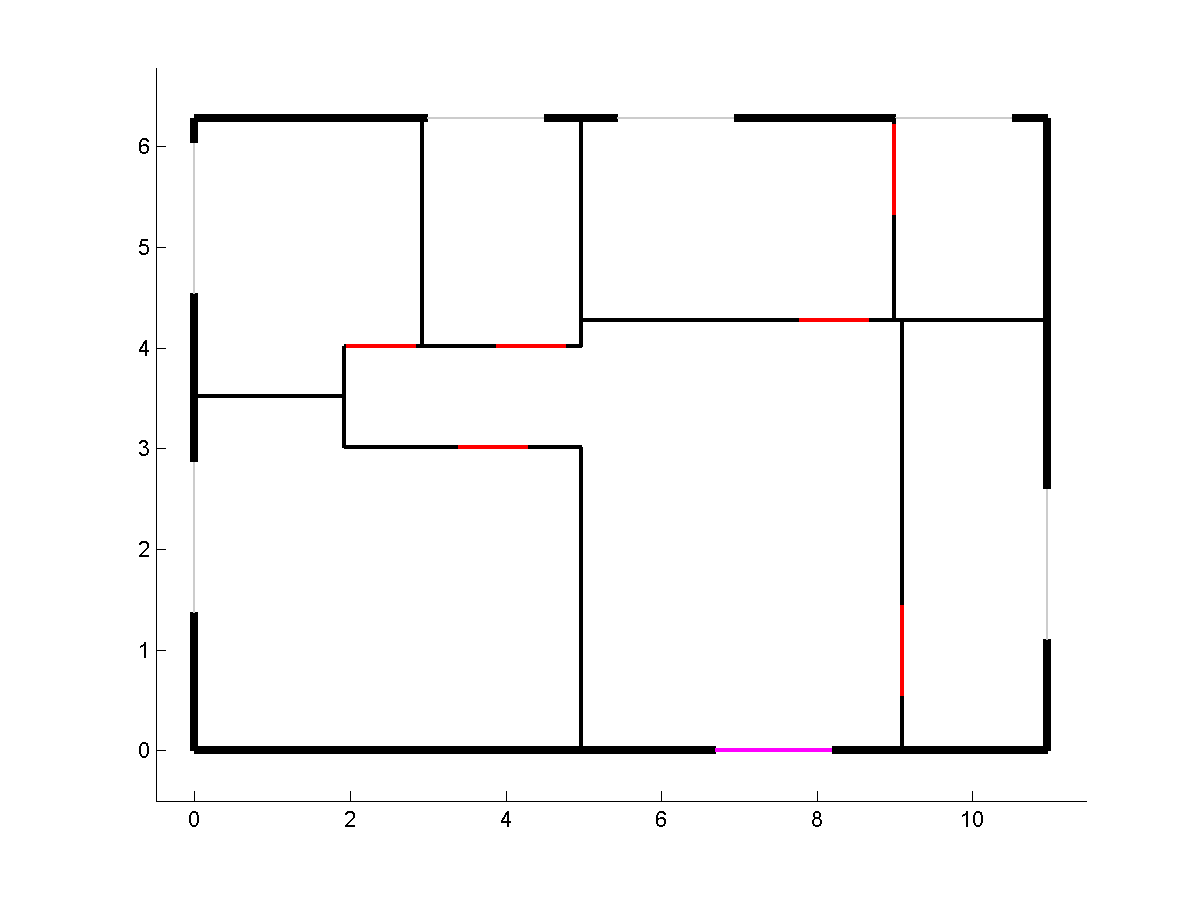} &  \includegraphics[scale=.25]{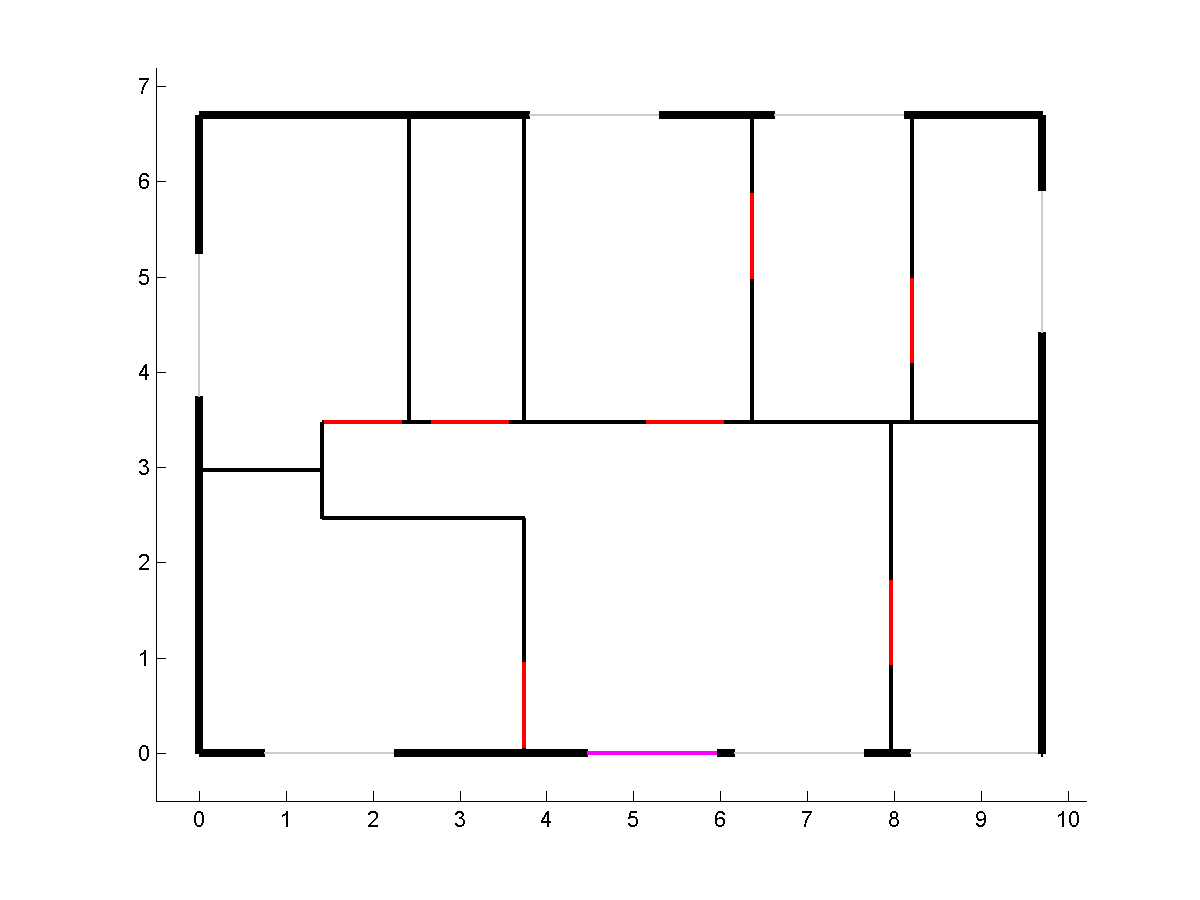} & \includegraphics[scale=.25]{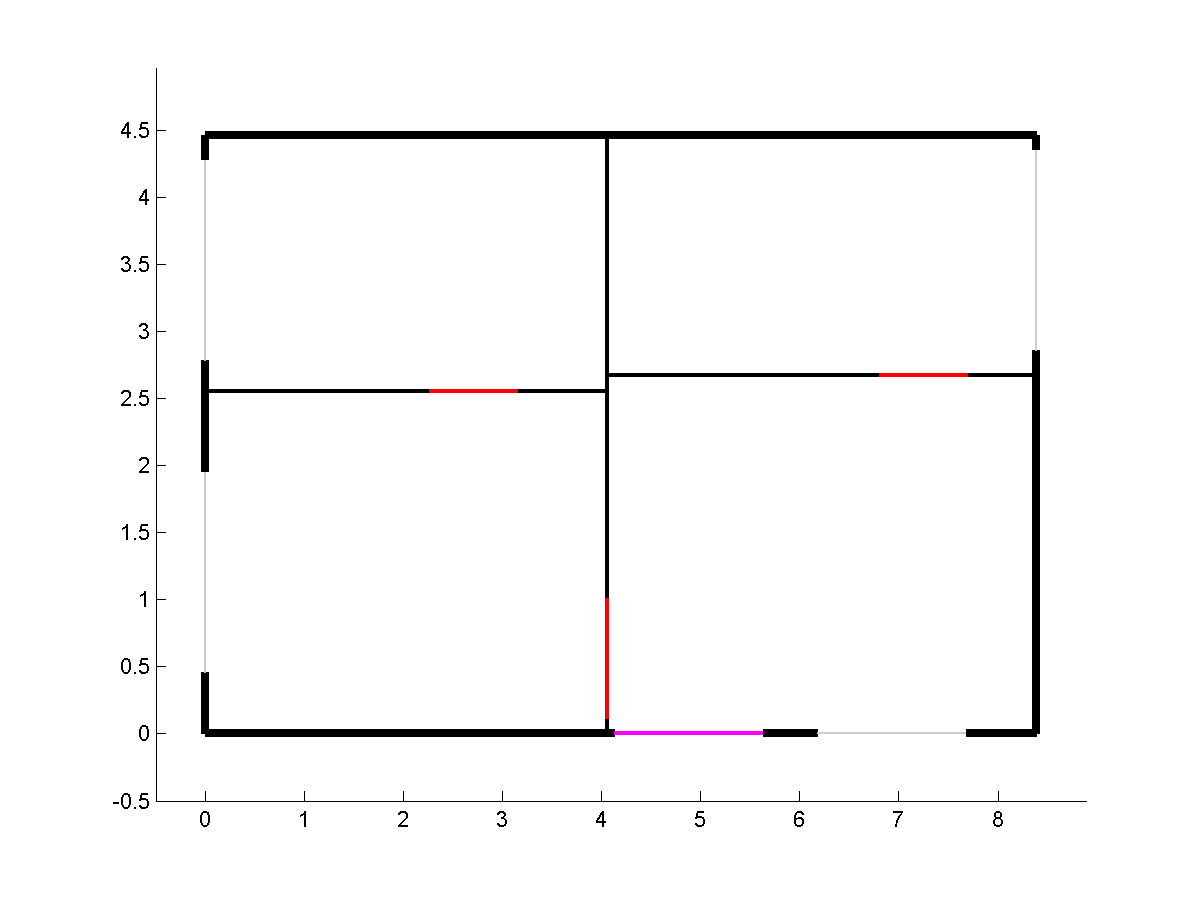} \\ 
\hline 
\includegraphics[scale=.25]{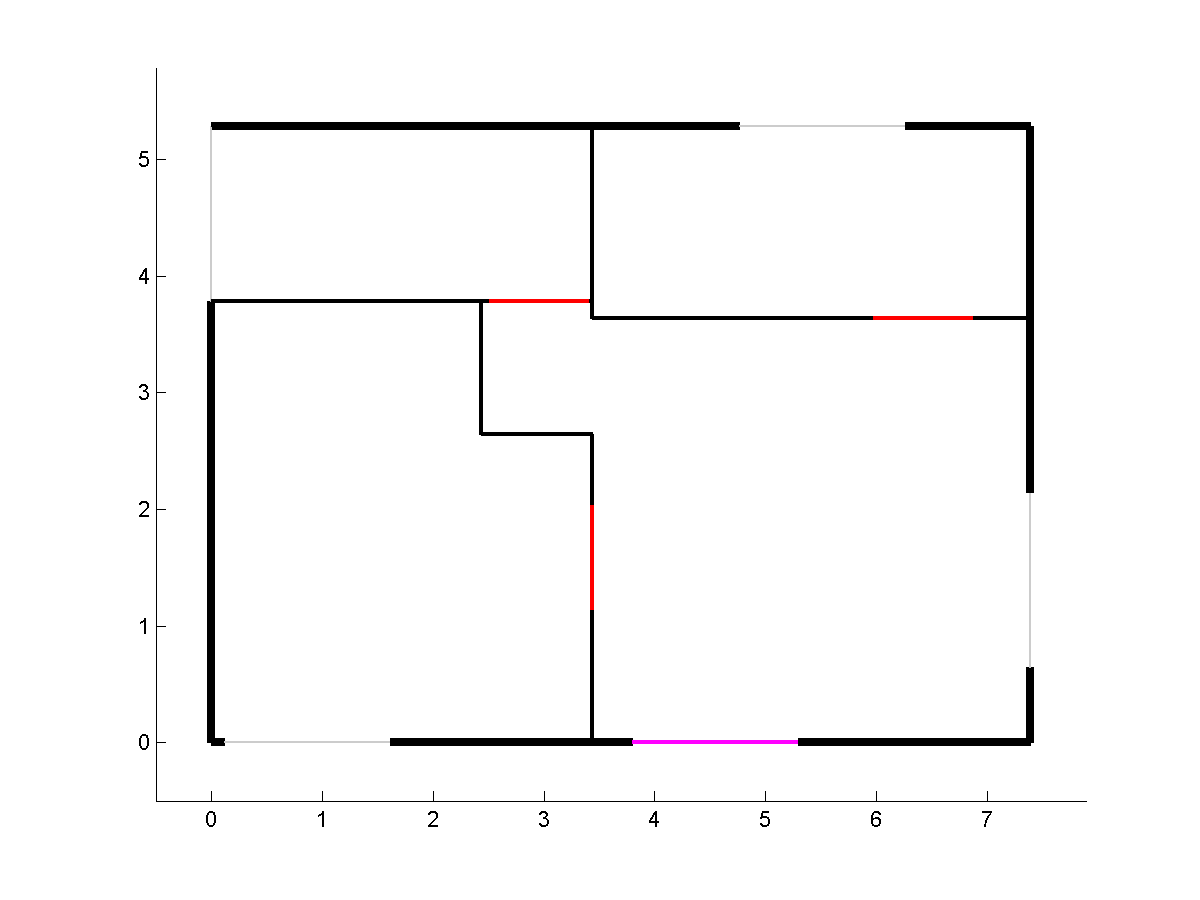} &  \includegraphics[scale=.25]{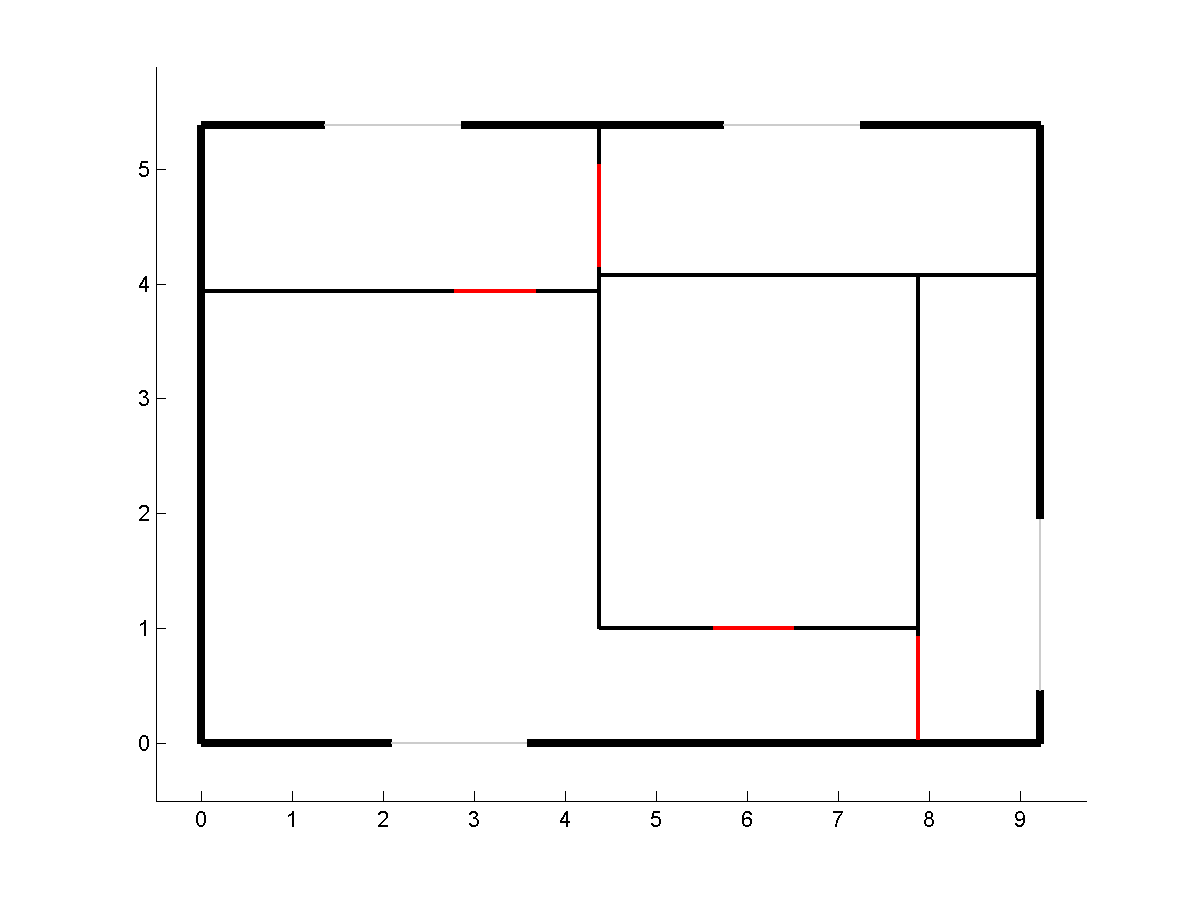} & \includegraphics[scale=.25]{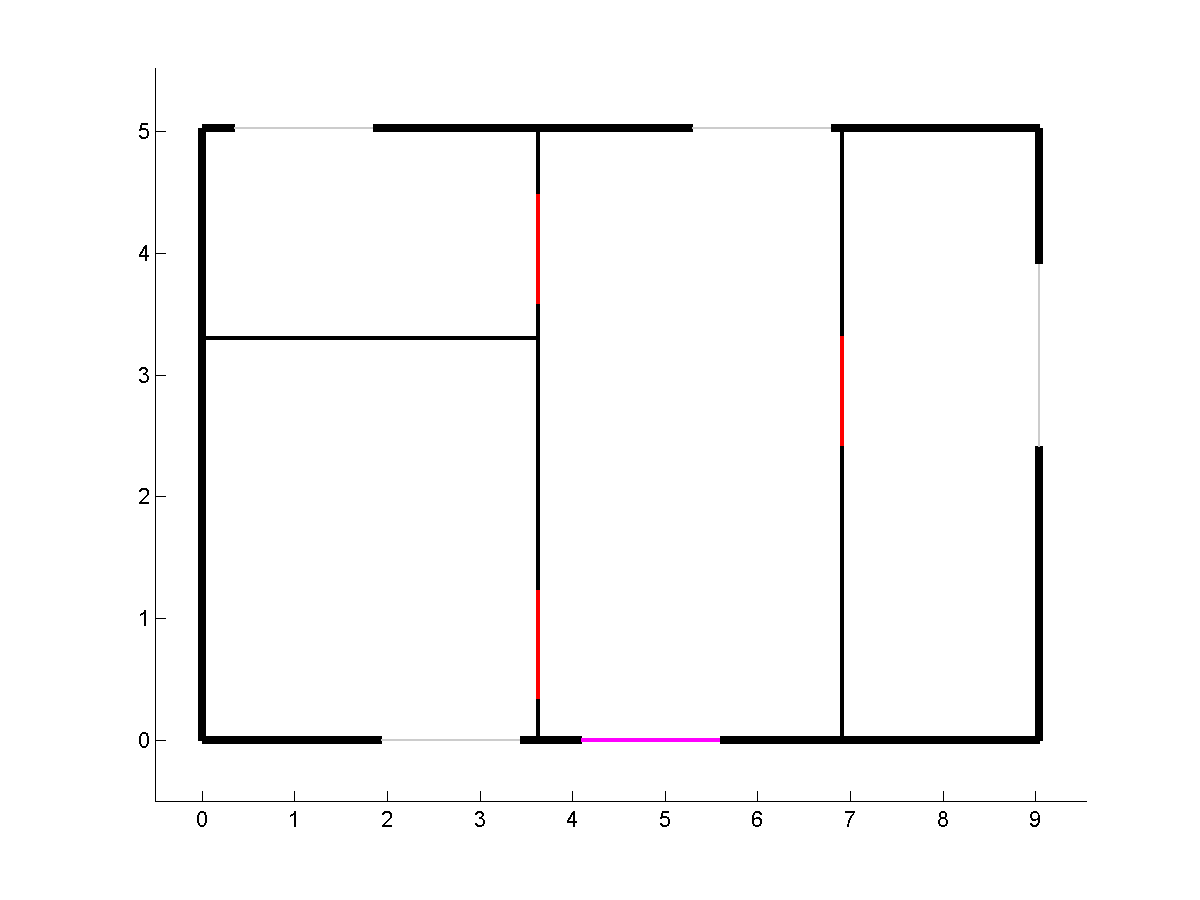} \\ 
\hline 
\end{tabular}

\end{document}